\documentclass[12pt,epsf]{article}
\input{amssym.def}
\input{amssym.tex}
\input{epsf}
\catcode`\@=11
\newcount\hour
\newcount\minute
\newtoks\amorpm
\hour=\time\divide\hour by60
\minute=\time{\multiply\hour by60 \global\advance\minute by-\hour}
\edef\standardtime{{\ifnum\hour<12 \global\amorpm={am}%
        \else\global\amorpm={pm}\advance\hour by-12 \fi
        \ifnum\hour=0 \hour=12 \fi
        \number\hour:\ifnum\minute<10 0\fi\number\minute\the\amorpm}}
\edef\militarytime{\number\hour:\ifnum\minute<10 0\fi\number\minute}
\def\draftlabel#1{{\@bsphack\if@filesw {\let\thepage\relax
   \xdef\@gtempa{\write\@auxout{\string
      \newlabel{#1}{{\@currentlabel}{\thepage}}}}}\@gtempa
   \if@nobreak \ifvmode\nobreak\fi\fi\fi\@esphack}
        \gdef\@eqnlabel{#1}}
\def\@eqnlabel{}
\def\@vacuum{}
\def\draftmarginnote#1{\marginpar{\raggedright\scriptsize\tt#1}}
\def\draft{\oddsidemargin -.2truein
        \def\@oddfoot{\sl preliminary draft \hfil
        \rm\thepage\hfil\sl\today\quad\militarytime}
        \let\@evenfoot\@oddfoot \overfullrule 3pt
        \let\label=\draftlabel
        \let\marginnote=\draftmarginnote
   \def\@eqnnum{(\theequation)\rlap{\kern\marginparsep\tt\@eqnlabel}%
\global\let\@eqnlabel\@vacuum}  }
\relax
\def\sqr#1#2{{\vcenter{\vbox{\hrule height.#2pt
        \hbox{\vrule width.#2pt height#1pt \kern#1pt
           \vrule width.#2pt}
        \hrule height.#2pt}}}}
\def\square{\mathchoice\sqr64\sqr64\sqr{2.1}3\sqr{1.5}3}
\def\lsim{{\displaystyle
{{\raise-8pt\hbox{$ <$}}
\atop{\raise5pt\hbox{$\sim$}}}}}
\def\gsim{{\displaystyle
{{\raise-8pt\hbox{$ >$}}
\atop{\raise5pt\hbox{$\sim$}}}}}
\def\slsim{{\displaystyle
{{\raise-8pt\hbox{$\scriptstyle <$}}
\atop{\raise5pt\hbox{$\scriptstyle \sim$}}}}}
\def\sgsim{{\displaystyle
{{\raise-8pt\hbox{$\scriptstyle  >$}}
\atop{\raise5pt\hbox{$\scriptstyle \sim$}}}}}
\newcommand{\sump}[0]{\sum_{(h,g)}\!{\raise 4pt \hbox{$'$}}\,}
\def\ifdd{\int_{\cal F}\frac{{\rm d}^2\tau}{\t_2^2}}
\def\na{\nabla}
\def\phl{\vphantom{l}}
\def\vep{\varepsilon}
\def\a{\alpha}
\def\b{\beta}
\def\g{\gamma}
\def\d{\delta}
\def\e{\epsilon}
\def\m{\mu}
\def\n{\nu}
\def\t{\tau}
\def\p{\pi}
\def\ps{\psi}
\def\r{\rho}

\def\s{\sigma}
\def\l{\lambda}
\def\k{\kappa}

\def\et{\eta}

\def\Ga{\Gamma}

\def\Fi{\Phi}
\def\O{\Omega}
\def\D{\Delta}
\def\bz{\bar{z}}
\def\bT{\bar{T}}

\def\bs{\bar{s}}

\def\bps{\bar{\psi}}

\def\Rr{{\cal R}}

\def\F{{\cal F}}

\def\cA{{\cal A}}

\def\bT{\bar{T}}

\def\Z{Z\!\!\! Z}

\def\pa{\partial}
\def\bpa{\bar{\partial}}

\def\ra{\rightarrow}

\def\ti{\times}

\def\bb{\bar{b}}
\def\ba{\bar{a}}

\def\un{\underline}

\def\nl{\newline}

\def\thefootnote{\fnsymbol{footnote}}
\def\be{\begin{equation}}
\def\ee{\end{equation}}
\def\bs{\begin{subequations}}
\def\es{\end{subequations}}
\def\ben{\begin{enumerate}}
\def\een{\end{enumerate}}
\def\ba{\begin{eqnarray}}
\def\ea{\end{eqnarray}}

\def\vs{\vskip}

\def\rd{{\rm d}}

\def\nl{\newline}
\def\ed{\end{document}}

\def\bibtem#1{\bibitem{#1} }

\def\sp{~~,~~}

\def\pe{\, .}
\def\co{\, ,}

\font\mybb=msbm10 at 12pt
\def\bb#1{\hbox{\mybb#1}}
\def\Z{\bb{Z}}
\def\R{\bb{R}}
\def\Z{\bb{Z}}
\def\R{\bb{R}}

\def\Ga{\Gamma}
\def\p{\pi}
\def\t{\tau}
\def\rd{{\rm d}}
\def\n{\nu}
\def\et{\eta}
\def\w{\wedge}

\def\ed{\end{document}}
\newtoks\@stequation
\def\subequations{\refstepcounter{equation}%
  \edef\@savedequation{\the\c@equation}%
  \@stequation=\expandafter{\theequation}%   %only want \theequation
  \edef\@savedtheequation{\the\@stequation}% % expanded once
  \edef\oldtheequation{\theequation}%
  \setcounter{equation}{0}%
  \def\theequation{\oldtheequation\alph{equation}}}

\def\endsubequations{\setcounter{equation}{\@savedequation}%
  \@stequation=\expandafter{\@savedtheequation}%
  \edef\theequation{\the\@stequation}\global\@ignoretrue
  \vspace*{-12pt} \\}
\def\thefootnote{\fnsymbol{footnote}}
\def\bea{\begin{eqnarray}}
\def\eea{\end{eqnarray}}
\def\be{\begin{equation}}
\def\ee{\end{equation}}
\def\bs{\begin{subequations}}
\def\es{\end{subequations}}
\newskip\humongous \humongous=0pt plus 1000pt minus 1000pt
\def\caja{\mathsurround=0pt}
\def\eqalign#1{\,\vcenter{\openup1\jot \caja
        \ialign{\strut \hfil$\displaystyle{##}$&$
        \displaystyle{{}##}$\hfil\crcr#1\crcr}}\,}
\newif\ifdtup

\def\un{\underline}

\def\thebibliography#1{%
\vskip 0.5cm \centerline{\bf References}
\list{%
[\arabic{enumi}]}{\settowidth\labelwidth{[#1]}
\leftmargin\labelwidth
\advance\leftmargin\labelsep
\usecounter{enumi}}
\def\newblock{\hskip .11em plus .33em minus .07em}
\sloppy\clubpenalty4000\widowpenalty4000
\sfcode`\.=1000\relax}

\renewcommand{\theequation}{\arabic{section}.\arabic{equation}}
\begin{document}

\begin{titlepage}
\begin{flushright}
hep-th/9906018\\
\end{flushright}
\begin{centering}
\vspace{.3in}
\boldmath
{\Large\bf Duality and Instantons in String Theory} \\
\unboldmath
\vspace{1.1 cm}
\large ELIAS KIRITSIS\footnote{e-mail address: kiritsis@physics.uch.gr}
\vskip 1cm
{\it Physics Department, P.O. Box 2208, University of Crete
}\\
{\it GR-71003 Heraklion, GREECE}\\
\medskip
\vspace{2.1cm}
{\bf Abstract}\\
\end{centering}
\vspace{.1in}

In these lecture notes  duality tests and instanton effects
in supersymmetric vacua of string theory are discussed.
 A broad overview of BPS-saturated terms in the effective actions is first given.
Their role in testing the consistency of duality conjectures
as well as discovering the rules of instanton calculus in string theory is discussed.
The example of heterotic/type-I duality is treated in detail.
Thresholds of $\F^4$ and $\Rr^4$ terms are used to test the duality as well
as to derive rules for calculated with D1-brane instantons.
We further consider the case of $\Rr^2$ couplings in N=4 ground-states.
Heterotic/type II duality is invoked to predict the heterotic NS5-brane instanton
corrections to the $\Rr^2$ threshold.
The $\Rr^4$ couplings of type-II string theory with maximal supersymmetry
are analysed and the D-instanton contributions are described 
Other applications and open problems are sketched.

\vspace{.9cm}
\begin{flushleft}
\vskip 2cm
May 1999 \\
\end{flushleft}
\vskip 2cm
\centerline{\tt Based on lectures given at the 1999 Trieste Spring School on String Theory}

\vskip 2cm

\end{titlepage}

\tableofcontents
\newpage
\setcounter{footnote}{0}
\renewcommand{\thefootnote}{\arabic{footnote}}

\setcounter{section}{0}

\section{Introduction}

Non-perturbative dualities have changed our way of thinking about 
string theory. They have also gave us the possibility to calculate non-perturbative
effects that we expected to be there but had no a priori way of setting up, let alone 
calculate.

There two basic issues that will be addressed in these lecture notes.
They are inextricably linked to one-another.
One is ``testing" duality conjectures. The second is putting them to work.

We will have to be clear in what we mean by ``testing" a duality conjecture.
In a theory where we do not have an a priori non-perturbative definition
a weak-strong coupling duality is a {\it definition} of the strongly coupled theory.
In general and for minimal supersymmetry it might not even be a complete non-perturbative
definition.
The issue becomes non-trivial if there a independent non-perturbative definition of the
theory. This however is not the case so far for supersymmetric theories since the only
other known non-perturbative definition, namely formulation on a lattice, breaks
supersymmetry. 
Thus, the only issue at stake in the case of duality is consistency of the definition.
Even in the case of field theory we know of examples where consistency alone poses
constraints in the non-perturbative definition of a theory.
It is for example well known that cluster decomposition and defining the non-perturbative
theory without including (smooth) monopoles are inconsistent.
Thus, what we are testing at best is the consistency of the set of rules that duality
uses to define the non-perturbative theory.

The consistency checks are stringent for effective couplings that have special properties.
We call them BPS-saturated effective couplings.
In a sense that will be made more precise later, supersymmetry constraints the form of 
their thresholds.
They are the most reliable tools in checking consistency of duality conjectures.

There have been  many qualitative checks of various non-perturbative
dualities, but so far quantitative checks are scarce.
In order to do a tractable quantitative test of a non-perturbative
duality
we need to carefully choose the quantity to be computed.
Since usually a weak coupling computation has to be compared with a
strong coupling one, one has to choose a quantity whose strong coupling
computation
can also be done at weak coupling.
Such quantities are very special and generally turn out to be terms in
the
effective action that obtain loop contributions from BPS states only.
They are also special from the supersymmetry point of view, since the
dependence of their couplings on moduli must satisfy certain
holomorphicity
or harmonicity  conditions.
Moreover, when supersymmetry commutes with the loop expansion, they get
perturbative corrections from a single order in perturbation theory.
Such terms also have special properties concerning instanton
corrections
to their effective couplings.
In particular, they obtain corrections only from instantons that
leave some part
of the original supersymmetry unbroken.
Sometimes, such terms are directly linked to anomalies.

The other role such effective couplings play is in that duality can be used to calculate
their non-perturbative corrections. One can then identify the non-perturbative effects
responsible for such corrections. For theories with more than N=2 supersymmetry
\footnote{We count the
supersymmetries using four-dimensional language (in units of four
supercharges).} such non-perturbative effects are due to instantons.
Instantons in string theory can be associated to Euclidean branes wrapping around an
appropriate compact manifold.
By studying such non-perturbative thresholds we can learn the rules of instanton
calculus.

While  solitons have been studied vigorously in the context of duality, 
the attention
paid to  instantons has been  lesser and more recent:
it includes work on the point-like D-instanton of type IIB
\cite{Po}--\cite{rt}, on the resolution of the type-IIA
conifold
singularity by  Euclidean 2-branes \cite{BBS}--\cite{O}, and on
non-perturbative effects associated with  Euclidean 5-branes
\cite{W}--\cite{6}. Here we will first look as an instructive example, 
at a  simpler case, that  of
Euclidean D-strings present in type-I $SO(32)$ string
theory \cite{bk2,ko,string97}:
these are physically less interesting, since they  are mapped
by strong/weak-coupling dualities
  to standard world-sheet instanton effects on the
type-IIB, respectively heterotic side. Our motivation is however
to gain a better understanding of the rules of semi-classical
D-instanton calculations, which  could prove useful in more interesting
contexts. 
We will further consider the case of $\Rr^2$ couplings in N=4 ground-states \cite{HM,6}.
In this case the relevant instanton corrections as we will argue are due
to the NS5-brane in the heterotic string \cite{HM}, or the D5-brane in the type-I
string. the threshold on the other hand is one-loop exact in the type-II 
dual.
Although we will go some way towards interpreting the threshold, a complete
instanton calculation is lacking here. 

Finally there is another area where duality and D-instanton corrections are 
at play. This is the case of the threshold corrections to eight-derivative terms in type-II vacua with maximal supersymmetry.
A representative eight-derivative term is the $\Rr^4$ term.
Here one can use T-duality, perturbative Dp-brane dynamics and eleven-dimensional input to 
successfully calculate the non-perturbative corrections.
This has been done down to six-dimensional compactifications, 
\cite{GG2},\cite{GV}-\cite{anto},\cite{kpar1}-\cite{eisen}
with a fairly good understanding of the D-instanton calculus, but not without some puzzles.  

The structure of these lectures is as follows:
In section two we provide a general discussion on the nature and properties of
BPS-saturated terms.
In section three we give a survey of known such terms in theories with varying amounts of
global or local symmetry.
In section four we discuss the anticipated structure of instantons in string theory as
well their similarities and differences with standard field theory instantons.
In section five we discuss in detail heterotic/type-I duality in various dimensions.
The two theories are compared in perturbation theory in nine-dimensions.
In eight dimensions there are D1-instanton corrections on the type-I side that 
are mapped by the duality to the perturbative heterotic contribution.
We use this to derive the instanton rules.
In section six we analyse $\Rr^2$ couplings in N=4 ground-states.
The threshold is one-loop exact in the type II string on K3$\times T^2$ and related via
duality to NS5-brane instanton corrections in the heterotic dual.
IN section seven we summarize the results on the $\Rr^4$-threshold in type-II 
ground-states with maximal supersymmetry.
Finally section eight contains a short summary as well a survey of open 
problems.

\section{BPS-saturated terms and Non-renormalization Theorems\label{bps}}
\setcounter{equation}{0}

Supersymmetry plays an important role in uncovering and testing the consistency
of duality symmetries.
It provides important constraints into the dynamics. Moreover, it has special
non-renormalization theorems that help in discerning some properties of the
strong coupling limit.

A central role is played by the BPS states. These are ``shorter than normal"
representations of the appropriate supersymmetry algebra\footnote{For a more
complete discussion, see \cite{book,lec}.}.
 Their existence is due to some of the supersymmetry operators
becoming ``null",
in which case they do not create new states.
The vanishing of some supercharges depends on the relation between the
mass of a multiplet and some central charges appearing in the
supersymmetry algebra.
These central charges depend on electric and magnetic charges of the
theory as well as on expectation values of scalars (moduli) that determine
various coupling constants.
In a sector with given charges, the BPS states are the lowest lying
states
and they saturate the so-called BPS bound which, for point-like states,
is of the form
\be
M\geq ~{\rm maximal~~eigenvalue~~of }\;\; Z
\,,\label{569}\ee
where $Z$ is the central charge matrix.

Massive BPS states appear in theories with extended supersymmetry.
BPS states behave in a very special way:

$\bullet$ At generic points in moduli space  they are absolutely
stable.
The reason is the dependence of their mass on conserved charges.
Charge and energy conservation prohibits their decay.
Consider as an example, the BPS mass formula
\be
M^2_{m,n}={|m+n\tau|^2\over \tau_2}\;\;,
\label{du2}
\ee
where $m,n$ are integer-valued conserved charges, and $\tau$ is a
complex modulus. This BPS formula is relevant for N=4, SU(2) gauge
theory, in a subspace of its moduli space.
Consider a BPS state with charges $(m_0,n_0)$, at rest, decaying into N
states
with charges $(m_i,n_i)$ and masses $M_i$, $i=1,2,\cdots,N$.
Charge conservation implies that $m_0=\sum_{i=1}^N m_i$,
$n_0=\sum_{i=1}^N n_i$.
The four-momenta of the  produced particles are $(\sqrt{M_i^2+\vec
p_i^2},\vec p_i)$ with $\sum_{i=1}^N \vec p_i=\vec 0$.
Conservation of energy implies
\be
M_{m_0,n_0}=\sum_{i=1}^N\sqrt{M_i^2+\vec p^2_i}\geq \sum_{i=1}^N
M_i\;\;.
\label{du1}\ee
Also in a given charge sector (m,n) the BPS bound implies that any mass
$M\geq M_{m,n}$, with $M_{m,n}$ given in (\ref{du2}).
Thus, from (\ref{du1}) we obtain
\be
M_{m_0,n_0}\geq \sum_{i=1}^N M_{m_i,n_i}\;\;,
\label{du3}
\ee
and the equality will hold if all  particles are BPS and are produced
at rest ($\vec p_i=\vec 0$).
Consider now the two-dimensional vectors $v_i=m_i+\tau n_i$ on the
complex $\tau$-plane, with length $||v_i||^2=|m_i+n_i\tau|^2$.
They satisfy $v_0=\sum_{i=1}^N v_i$.
Repeated application of the triangle inequality implies
\be
||v_0||\leq \sum_{i=1}^N ||v_i||\;\;.
\label{du4}
\ee
This is incompatible with energy conservation (\ref{du3}) unless
all vectors $v_i$ are parallel. This will happen only if $\tau$ is
real.\footnote{It can also happen when the charges are integer multiples
of the same charge. In that case the composite states are simple bound states
of some ``fundamental" states. We will not worry further about this
possibility.}
For energy conservation it should also be a rational number.
On the other hand, due to the SL(2,$\Z$) invariance of (\ref{du2}), the
inequivalent choices for $\tau$ are in the SL(2,$\Z)$ fundamental
domain and
$\tau$ is never real there. In fact, real rational values of $\tau$ are
mapped by SL(2,$\Z)$ to $\tau_2=\infty$, and since $\tau_2$ is the
inverse of the coupling constant, this corresponds to the degenerate
case of zero coupling.
Consequently, for $\tau_2$ finite, in the fundamental domain, the BPS
states of this theory are absolutely stable. This is always true in
theories
with more than eight conserved supercharges (corresponding to N$>2$
supersymmetry
in four dimensions).
In cases corresponding to theories with 8 supercharges, there are
regions in the moduli space, where BPS states, stable at weak coupling,
can decay at strong coupling. However, there is always a large region
around weak coupling where they are stable.

$\bullet$ Their mass-formula is supposed to be exact if one uses
renormalized values for the charges and moduli.
The argument is that quantum corrections would spoil the relation of
mass and charges, if we assume unbroken SUSY at the quantum level.
This would give incompatibilities  with the dimension of their
representations.
Of course this argument seems to have a loophole: a specific set of BPS
multiplets can combine into a long one. In that case, the above
argument
does not prohibit corrections.
Thus, we have to count BPS states modulo long supermultiplets.
This is precisely what helicity supertrace formulae do for us \cite{book,lec}.
Even in the case of N=1 supersymmetry there is an analog of BPS states,
namely the massless states.

Thus, the presence of BPS states can be calculated at weak coupling and can be
trusted in several cases at strong coupling.
Their mass-formula is valid beyond perturbation theory, although both sides can
obtain
non-trivial quantum corrections in N=2 supersymmetric theories.
There are no such corrections in N$\geq 3$ supersymmetric theories.

There is another interesting issue in theories with supersymmetry: special
terms in the effective action, protected by supersymmetry.
We will generically call such terms ``BPS-saturated terms" for reasons that
will become obvious in the sequel.
We cannot at the moment give a rigorous definition of such terms for a generic
supersymmetric theory, due to the lack of off-shell formulation.
However, for theories with an off-shell formulation the situation is better
and all BPS-saturated terms are known.

We will try here before we embark in a detailed discussion of various cases
to give some generic features of such terms.

(a)  Supersymmetry constraints their (moduli-dependent) coefficients
to have a special structure. The simplest situation is complex holomorphicity
but there several other cases less common where one would have special
conditions associated with quaternions, as well as non-compact groups of the
O(p,q) type , etc.
We will generically call such constraints
``holomorphicity" constraints.

(b)  In cases where there is a superfield formulation, such terms are
(chiral) integrals over parts of superspace. This is at the root of the
holomorphicity conditions of the previous item.

(c)  Such terms satisfy special non-renormalization theorems.
It should be stressed here and it would be seen explicitly later that some of
the non-renormalization theorems depend crucially on the perturbation theory
setup.

Before we embark further into a discussion of the non-renormalization theorems
we should stress in advance that they are generically only valid for the
Wilsonian
effective action. The reason is that many non-renormalization theorems are
violated
due to IR divergences.
There are several examples known, we will mention here the case of N=1
supersymmetry:
In the presence of massless of massless contributions a quantum correction
to the K\"ahler potential (not protected by a non-renormalization theorem) can
be indistinguishable from a correction to the superpotential (not renormalized
in perturbation theory), \cite{west}.
Thus, from now on we will be always discussing the Wilsonian effective action.

To continue further, we would like here to separate two cases.

\begin{itemize}
\item Absolute non-renormalization theorems. Such theorems state that a given
term
in the effective action of a supersymmetric theory does not get renormalized.
This should be  valid both for perturbative and non-perturbative corrections. A
typical example of this is the case of two derivative terms in
the effective action of an N=4 supersymmetric theory (with or without gravity).

\item Partial (or perturbative)  non-renormalization theorems.
Such theorems usually claim the absence of perturbative corrections for a given
effective coupling, or that the only corrections appear in a few orders only
in perturbation theory.
Typically this happens at one-loop order but we also know of cases where
renormalization can occur at a single, arbitrarily-high, loop order.
It is also common in the case of one-loop corrections only that the appropriate
effective couplings are related to an anomaly via supersymmetry.
The appropriate Adler-Bardeen type theorem for the anomaly guarantees the
absence of higher loop corrections, provided the perturbation theory is set up
to respect
supersymmetry.
An example of such a situation is the case of two-derivative couplings in a
theory with N=2 supersymmetry (global or local).

\end{itemize}

One should note a potentially very interesting generalization of supersymmetric non-renormalization theorems: their analogues in the case where supersymmetry in softly broken (in field theory) or spontaneously broken (in supergravity or string theory).
Although, there are some results in this direction mostly in field theory \cite{gg}, 
I believe that much more needs to be done.
Moreover this subject is of crucial importance in any unified theory that uses supersymmetry as a solution to the hierarchy problem.

It should be stressed here that the way perturbation theory is setup is crucial
for the applicability of such partial non-renormalization theorems.
Many of the non-perturbative string dualities amount simply to different
perturbative expansions of the same underlying theory. As we will see in more
detail later on
the appropriate partial non-renormalization theorems for the same effective
coupling
are different in dual versions. In many cases this can be crucial in obtaining
the exact result.

There are several examples that illustrate the general discussion above.
We will mention some commonly known ones.

1. Heterotic string theory on $T^6$ is dual to type II string theory on
K3$\times T^2$.
The $\Rr^2$ effective coupling, has only a one-loop contribution on the type II
side.
On the heterotic side apart from the tree-level contribution there are no
other perturbative corrections. There are however non-perturbative corrections
due to five-brane
instantons \cite{HM,6}.

2. Heterotic string theory on   K3$\times T^2$ is dual to type IIA on a
Calabi-Yau (CY)
manifold that is a K3-fibration.
All two derivative effective couplings are tree-level only on the type-II side.
On the other hand they obtain tree-level, one-loop as well as non-perturbative
corrections on the heterotic side.

It can sometimes happen that a given perturbative expansion does not commute
with supersymmetry.
This is the case generically in type-I string theory. One way to see this is to
note that the leading correction to the $trF^4$ term in ten dimensions comes
from the disk
diagrams while the $B\wedge tr F^4$ term obtains a contribution from one-loop
only (Green-Schwarz anomaly-canceling term).
The two terms are however related by supersymmetry \cite{roo}.

There is another general property that is shared by BPS-saturated terms:
The quantum corrections to their effective couplings can be associated to BPS
states.
There are two concrete aspects of the statement above:

$\bullet$ Their one-loop contributions are due to (perturbative) BPS multiplets
only.\footnote{This was observed in \cite{hm2} for N=2 two-derivative
couplings,
and in \cite{BaKi,lec} for N=4 four-derivative couplings and N=8
eight-derivative couplings.} The way this works out is that the appropriate
one-loop diagrams come out
proportional to helicity supertraces.
The helicity supertrace is a supertrace in a given supersymmetry representation
of casimirs of the little group of the Lorentz group \cite{str}.
In four dimensions, this is a supertrace of the helicity $\lambda$
to an arbitrary
even power (by CPT all odd powers vanish):
$$
B_{2n}={\rm Str}~[\lambda ^{2n}]
$$

The helicity supertraces are essentially indices to which only
short  BPS multiplets contribute \cite{BaKi,lec}.
It immediately follows that such one-loop
contributions
are due to BPS states only.
The appropriate helicity supertraces count essentially the numbers of
``unpaired" BPS multiplets. It is only these that are protected from
renormalization
and can provide reliable information in strong coupling regions.
In fact, calling the helicity supertraces indices is more
than an analogy.
They thus provide the minimal information about IR-sensitive data.
In particular they do not depend on the moduli.
Unpaired BPS states in lower dimensions are intimately
connected with the chiral asymmetry (conventional index) of the
ten-dimensional (or eleven-dimensional)
theory.
It is well known that the ten-dimensional elliptic genus is the stringy
generalization of the Dirac index \cite{Windey,ellwit}.
Projecting the elliptic genus on
physical states
in ten dimensions gives precisely the massless states, responsible
for anomalies.
In lower dimensions, BPS states are determined uniquely by the elliptic
genus,
as well as the compact manifold data.

We will describe here a bit more the properties of helicity supertraces. Further
information
and detailed formulae can be found in the appendix of \cite{book}.

N=2 supersymmetry. Here we have only one kind of a BPS multiplet the 1/2
multiplet
(preserving half of the original N=2 supersymmetry).
The trace $B_2$ is non-zero for the 1/2 multiplet but zero from the long
multiplets.
The long multiplets on the other hand contribute to $B_4$.

N=4 supersymmetry. Here we have two types of BPS multiplets, the 1/2 multiplets
and
the 1/4 multiplets. For all multiplets $B_2=0$. $B_4$ is non-zero for 1/2
multiplets only.
$B_6$ is non-zero for 1/2 and 1/4 multiplets only. Long massive multiplets
start contributing
only to $B_8$.
A similar stratification appears also in the case of maximal N=8 supersymmetry
\cite{book}.

Thus, there is a single ``index" in the case of an N=2 supersymmetric theory,
and it governs
the one-loop corrections to the two-derivative action, in standard perturbation
theory.
In the N=4 case there are two distinct indices. $B_4$ controls one-loop
corrections to several
four-derivative effective couplings of the $tr F^4$, $tr \F^2tr \Rr^2$ and $tr
\Rr^4$ type.
$B_6$ seems to be associated to  some six derivative couplings like $tr \F^6$.

$\bullet$ BPS-saturated couplings may receive also instanton corrections. The
instantons however
that contribute, parallel in a sense the BPS states that contribute in
perturbation theory.
They must preserve the same amount of supersymmetry.
Since in string theory instantons are associated with Euclidean solitonic
branes wrapped on
compact manifolds, it is straightforward in most situations to classify
possible instantons that
contribute to BPS-saturated terms. We will see explicit examples in subsequent
sections.

Let us summarize here the generic characteristics of BPS-saturated couplings.

(1) They obtain perturbative corrections from BPS states only.

(2) The perturbative corrections appear at a single order of perturbation
theory, usually
at one-loop.

(3) They satisfy "holomorphicity constraints".

(4) They contain simple information about massless singularities.

(5) They obtain instanton corrections from "BPS-instantons" (instanton
configurations that
preserve some fraction of the original supersymmetry).

(6) If there exists an off-shell formulation they can be easily constructed.

Here we would like to remind the reader of a few facts about 
string perturbation theory.
In particular we focus on heterotic
 and type-II perturbation theory. Almost nothing is known for type-I 
 perturbation theory beyond one loop.
 
There are many subtleties in calculating higher-loop contributions
that arise from the presence of supermoduli.
There is no rigorous general setup so far, but several facts are known.
As discussed in \cite{pert} there are several prescriptions for
handling the
supermoduli. They differ by total derivatives on moduli space.
Such total derivatives can sometimes obtain contributions from the
boundaries of moduli space where the Riemann surface degenerates or
vertex operator
insertions collide.
Thus, different prescriptions differ by contact terms.
In \cite{cato} it was shown that such ambiguities eventually reduce to
tadpoles of massless fields at lower orders in perturbation theory.
The issue of supersymmetry is also the subject of such ambiguities.
It is claimed \cite{pert,cato} that in a class of prescriptions
$N\geq 1$ supersymmetry is respected genus by genus provided
there are no disturbing tadpoles at tree level and one loop.
The only exception to this is the case of an anomalous $U(1)$ in $N=1$
supersymmetric ground-states. In this case there  is a
non-zero D-term at one loop \cite{dterm}.

To conclude, if all (multi) tadpoles vanish at one loop and we use the
appropriate prescription for higher loops, we expect supersymmetry to
be valid order by order in perturbation theory.
It is to be remembered, however, that the above statements apply
on-shell.
Sometimes there can be terms in the effective action that vanish
on-shell,
violate the standard lore above, but are required by non-perturbative
dualities. An example was given in \cite{6}.

\section{Survey of BPS-saturated terms\label{survey}}
\setcounter{equation}{0}

In this section we will describe the known BPS-saturated terms for a given
amount of
supersymmetry. We use {\it four-dimensional language for the
supersymmetry},
which can eventually be translated to various
other dimensions bigger than four. For example N=2 four-dimensional theories
are related by
toroidal compactification to N=1 six-dimensional theories and so on.

\subsection{N=1 Supersymmetry\label{n=1}}

N=1 supergravity in four dimensions contains the supergravity multiplet (it
contains the
metric and a gravitino)
vector multiplets (each contains a vector and a gaugino) and chiral multiplets
(each contains a
complex scalar and a Weyl fermion).

The "critical" dimension is four, in the sense that we cannot have an N=1
theory in more than
four dimensions.

The full two-derivative effective action is determined by three
functions\footnote{A slightly
more extended description can be found in the appendix of \cite{book,lw}. For a
detailed exposition
as well as a discussion of the non-renormalization theorems \cite{n=1} the
reader is urged
to look in \cite{lk}.}

(a) The K\"ahler potential $K(z^i,\bar z^i)$. It is an arbitrary real function
of the complex
scalars $z^i$ of the chiral multiplets. It determines, among other things,  the
kinetic terms
of chiral multiplets (matter) via the K\"ahler metric $g_{i\bar
j}=\pa_{i}\pa_{\bar j}K$.

(b) The Superpotential $W(z^i)$: It determines the part of the scalar potential
associated to
the F-terms as $V_{F}\sim g^{i\bar j}\pa_{i}W\pa_{\bar j}\bar W$.
Supersymmetry constraints $W$ to be a holomorphic function of
the chiral multiplets. Moreover, it should have charge 2 under the U(1)
R-symmetry that
transforms the superspace coordinates $\theta \to e^{ia}\theta$ and the chiral
superfields
$Z^i\to e^{-ia}Z^i$.
The reason is that the superpotential is integrated over half of the superspace
as $\int
d^2\theta ~W$.
It is thus a "chiral" density. It turns out that in all cases where and
off-shell superfield
formulation exists all BPS-saturated terms are chiral densities.
Both in string theory (supergravity) and in the global supersymmetric limit
(decoupling of
gravity) it is not renormalized in perturbation theory.
In string theory the argument is based on the holomorphicity \cite{ds}.
The string coupling constant (dilaton), is assembled with the axion (dual of
the antisymmetric
tensor in four dimensions) into the complex chiral $S$ field. The Peccei-Quinn
symmetry
associated to translations of the axion is valid in string perturbation theory,
but it is broken
by non-perturbative effects since it is anomalous. In perturbation theory,
corrections to
$W$ are multiplied by powers of the coupling constant, $(Im ~S)^{-n}$. However
such corrections
must be holomorphic and thus proportional to $S^{-n}$. This however breaks the
Peccei-Quinn
symmetry in perturbation theory. Thus, no such perturbative corrections can
appear.
Beyond perturbation theory, instanton effects break the Peccei-Quinn symmetry
to some discrete
subgroup and exponential holomorphic corrections are allowed \cite{W}.
In the global limit a similar argument works.

(c) The Wilsonian gauge couplings $f_a(Z^i)$. They are also holomorphic. The
index $a$ labels different
simple or U(1) components of the gauge group. They are also chiral densities
since they appear
through $\int d^2\theta f_{a}(Z^i){\cal W}^{\a}{\cal W}_{\a}$ in the effective
action, where
${\cal W}^{\a}$ is the spinorial vector superfield.
They can obtain corrections only to one-loop in perturbation theory \cite{lk}.
The argument is similar to
that about the superpotential, the anomaly here allowing also a one-loop
contribution
We should stress here though that the physical effective gauge couplings have
corrections to any
order in perturbation theory. This is due to the fact that the physical matter
fields have a
wave function renormalization coming from, the K\"ahler potential which is not
protected from renormalization.

(d) In string theory, there is a possibility of ``anomalous" U(1) factors of
the gauge group
\cite{ds1}.
The term anomalous here is strictly speaking a misnomer. The particular U(1)
factor in question
will have a non-zero sum of charges for the massless fields. Under normal
circumstances
it would have been anomalous. In string theory, this anomaly is canceled by an
anomalous
transformation law of the antisymmetric tensor (or an axion in four dimensions). 
We will have to distinguish two distinct cases.

In {\it heterotic} perturbation theory the anomaly-cancellation
mechanism is a
compactification descendant of the Green-Schwarz anomaly cancellation in ten
dimensions.
The appropriate coupling is $B_{\m\n}F^{\mu\nu}$ which appears at one-loop in
the heterotic string.
Thus it is the standard antisymmetric tensor that cancels the anomaly.
There is a D-term contribution to the potential. It contains a one-loop
contribution that was
calculated in \cite{dterm}. Moreover supersymmetry implies that there should be
a two-loop contact
term. This was verified by an explicit calculation \cite{2loop}.
Such non-trivial contributions appear since the supersymmetric partner of the
antisymmetric tensor (axion) is the dilaton that controls the string perturbative
expansion.
The states that contribute at one loop are the charged
massless particles.
In this sense this is an anomaly. As mentioned earlier such states are the
closest analogue
of BPS states of N=1 supersymmetry (they have half the degrees of freedom
compared to massive
states).
Moreover, upon toroidal
compactification to three or two dimensions, they become bona-fide BPS states.
The two-dimensional case was analyzed in \cite{lerche}.

In {\it type}-II and {\it type}-I perturbation theory the situation is different \cite{ib}.
The axions that are responsible for canceling the U(1) anomaly come from the 
RR sector and they are usually more than one.
In orbifold constructions they belong to the twisted sector.
Since their scalar partners do not coincide with the string coupling constant
the D-term potential appears here at tree level.

(e) There is a series $I_g$ of higher derivative terms which are F-terms and 
involve chiral projectors of vector
superfields \cite{n=1t}. They obtain a perturbative contribution only at g-loops.

\subsection{N=2 Supersymmetry\label{n=2}}
\renewcommand{\theequation}{\thesubsection.\arabic{equation}}
\setcounter{equation}{0}

N=2 supersymmetry has critical dimension six.
The relevant massless multiplets in six dimensions are the supergravity
multiplet
(graviton, second rank tensor, a scalar a gravitino and a spin-half fermion),
the vector
multiplet (a vector and a gaugino), and the hypermultiplet (a fermion and four
scalars).
The vector multiplets contain no scalars in six dimensions and as such have a unique coupling to gravity.
This is not the case with hypermultiplets that have a non-trivial $\s$-model structure.
This structure persists unchanged in four dimensions, and we will discuss it below.

In four dimensions, the supergravity multiplet contains the metric a
graviphoton and two
gravitini. The vector multiplet contains a vector, two gaugini and a complex
scalar while the
hypermultiplet is the same as in six dimensions.

We will describe briefly the structure of the effective supergravity
 theory in four dimensions \cite{n=2}. The interested reader can consult
\cite{n=3} for further
 information.
Picking the gauge group to be abelian is without loss of generality since
any non-abelian gauge group can be broken to the maximal abelian
subgroup
by giving expectation values to the scalar partners of the abelian gauge
bosons.
We will denote the graviphoton by $A^0_{\m}$, the rest of the gauge
bosons
by $A^i_{\mu}$, $i=1,2,\dots,N_V$, and the scalar partners of
$A^i_{\m}$
as $T^i,\bar T^i$.
Although the graviphoton does does not have a scalar partner, it is
convenient
to introduce one.
The theory  has a scaling symmetry, which allows us to set this scalar
equal to one where K is the K\"ahler potential.
We will introduce the complex coordinates $Z^I$,
$I=0,1,2,\dots,N_V$,
which will parametrize the vector moduli space (VMS),~${\cal M}_V$.
The  $4 N_H$ scalars of the generically massless hypermultiplets parametrize
the
hypermultiplet moduli space ${\cal M}_H$ and supersymmetry requires
this to be a quaternionic manifold\footnote{In the global supersymmetry limit
in which gravity decouples, $M_{\rm Planck}\to \infty$, the geometry of the
hypermultiplet space is that of a hyperk\"ahler manifold.}.
The geometry of the full scalar manifold is that of a product, ${\cal
M}_V\times {\cal M}_H$.

N=2 supersymmetry implies that the VMS is not just a K\"ahler manifold,
but that it satisfies what is known as special geometry.
Special geometry eventually leads to the property that the full
action
of N=2 supergravity (we exclude hypermultiplets for the moment) can
be written
in terms of one function, which is holomorphic in the VMS coordinates.
This function, which we will denote by $F(Z^I)$, is called the
prepotential.
It must be a homogeneous function of the coordinates of degree 2:
$Z^I~F_I=2$, where $F_I={\pa F\over \pa Z^I}$.
For example, the K\"ahler potential is
\be
K=-\log\left[i(\bar Z^I~F_I-Z^{I}\bar F_I)\right]
\,,\label{442}\ee
which determines the metric $G_{I\bar J}=\pa_I\pa_{\bar J} K$ of
the kinetic terms of the scalars.
We can  fix the scaling freedom by setting $Z^0=1$, and then
$T^i=Z^i/Z^0$ are the physical moduli.
The K\"ahler potential becomes
\be
K=-\log\left[2\left(f(T^i)+\bar f(\bar T^i)\right)-(T^i-\bar
T^i)(f_{i}-\bar f_i)\right]
\,,\label{443}\ee
where $f(T^i)=-iF(Z^0=1,Z^i=T^i)$.
The K\"ahler metric $G_{i\bar j}$ has the following property
\be
R_{i\bar j k\bar l}=G_{i\bar j}G_{k\bar l}+G_{i\bar l}G_{k\bar j}
-e^{-2K}W_{ikm}G^{m\bar m}\bar W_{\bar m\bar j\bar l}
\,,\label{444}\ee
where $W_{ijk}=\pa_i\pa_j\pa_k f$.
Since there is no potential, the only part of the bosonic
action left
to be specified is the kinetic terms for the vectors:
\be
{\cal L}^{\rm vectors}=-{1\over
4}\Xi_{IJ}F^I_{\m\n}F^{J,\m\n}-{\theta_{IJ}\over 4}F^I_{\mu\n}\tilde
F^{J,\m\n}
\,,\label{451} \ee
 where
\be
\Xi_{IJ}={i\over 4}[N_{IJ}-\bar
N_{IJ}]\;\;\;,\;\;\;\theta_{IJ}={1\over 4}
[N_{IJ}+\bar N_{IJ}]
\,,\label{452}\ee
\be
N_{IJ}=\bar F_{IJ}+2i{{\rm Im} ~F_{IK}~{\rm Im}~ F_{JL}Z^KZ^L\over
{\rm Im}~F_{MN}Z^MZ^N}
\,.\label{453}\ee
Here we see that the gauge couplings, unlike the N=1 case, are not
harmonic functions of the moduli.

The self-interactions of massless hypermultiplets are described by a $\s$-model
on a quaternionic manifold (hyperk\"ahler in the global case)
A quaternionic manifold must satisfy:

(1) It must have three complex structures $J^i$, $i=1,2,3$ satisfying the
quaternion algebra
$$ J^i~ J^j=-1\delta^{ij}+\epsilon^{ijk}J^k
$$
with respect to which the metric is hermitian. The dimension of the manifold is
 4m,
$m\in \Z$.
The three complex structures guarantee the the existence of an SU(2)-valued
hyperk\"ahler two-form K.

(2) There exists a principal SU(2) bundle over the manifold, with connection
$\omega$
such that the form K is closed with respect to $\omega$
$$ \nabla K=dK+[\omega,K]=0$$

(3) The connection $\omega$ has a curvature that is proportional to the
hyperk\"ahler form
$$ d\omega+[\omega,\omega]=\lambda ~K
 $$
 where $\lambda$ is a real number.
 When $\lambda=0$ the the manifold is hyperk\"ahler.
 Thus the holonomy of a quaternionic manifold is of the form $SU(2)\otimes H$
while for a
 hyperk\"ahler manifold $H$, with $H\subset Sp(2m,\R)$.
The existence of the SU(2) structure is natural for the hypermultiplets since
SU(2)
is the non-abelian part of the N=2 R-symmetry that acts inside the
hypermultiplets.
The scalars transform as a pair of spinors under SU(2).
When the hypermultiplets transform under the gauge group, then the quaternionic
manifold has
appropriate isometries (gauging).
More information can be found in \cite{n=3}.

Thus, supersymmetry implies (a) that all two-derivative couplings in the vector
multiplets
are determined by a holomorphic function of the moduli, the prepotential.
Here the symmetry is $U(1)\to$ complex holomorphicity
(b) All two derivative couplings of the hypermultiplets are determined by
quaternionic geometry
$\to$ SU(2) ``holomorphicity".

In N=2 supersymmetry all two-derivative couplings are of the BPS-saturated
type.
Their precise non-renormalization properties though depend on the perturbation
theory
setup.

$\bullet$ \un{Global N=2 supersymmetry}. It can be obtained by taking the
$M_{P}\to \infty$ limit of
the locally supersymmetric case.
Here the holomorphic prepotential that governs the vector multiplet moduli
space
obtains quantum corrections only at one-loop in perturbation theory.
Beyond the perturbative expansion it obtains instanton corrections.
On the other hand the massless hypermultiplet geometry does not have any
perturbative or
non-perturbative corrections. In the quantum theory the only thing that can
change are the
points where various Higgs branches intersect themselves or the Coulomb branch,
\cite{as}.
The argument of \cite{as} for this non-renormalization is supersymmetry and
is an adaptation of a similar argument   valid in heterotic string theory to be
discussed below. In this context the crucial constraint imposed by
supersymmetry is
that the geometry of the vector moduli space is independent of the geometry of
the hypermoduli
space.
Put otherwise, the only couplings between vector and hypermultiplets are those
dictated
by the gauge symmetry.

$\bullet$ \un{Local N=2 supersymmetry}. Here we must distinguish three
different types of
perturbation theory.

(a) \un{Type-II perturbation theory}. This is relevant for type II ground-states
with (1,1)
four-dimensional supersymmetry. One of the supersymmetries comes from the left
moving sector
while the other from the right moving one.
A typical class of examples much studied is type IIA,B theory compactified on
CY threefold.
The type-IIA compactification gives an effective theory with $N_V=h_{1,1}$
vector multiplets
and $N_{H}=h_{1,2}+1$ neutral hypermultiplets (see for example \cite{book}).
In the type-IIB compactification the roles of $h_{1,1}$ and $h_{1,2}$ are
interchanged.
On the other hand such (1,1) ground-states need not be left-right symmetric.
What is an important feature of these ground-states is that the dilaton that
determines the
string coupling constant belongs to a hypermultiplet.
This has far reaching consequences for the perturbative expansion.
If this fact is combined with the supersymmetric constraint of the absence of
neutral couplings
between vectors multiplets and hypermultiplets, then we conclude:
in type-II (1,1) ground-states the tree-level prepotential is non-perturbatively
exact, while
the hypermultiplet geometry obtains corrections both in perturbation theory and
non-perturbatively.
Thus, the exact prepotential in type-IIA ground-states can be obtained by a
tree-level
calculation, in the appropriate $\s$-model. It describes the geometry of
K\"ahler structure
of the CY manifold.
Non-trivial $\s$-model instantons render this calculation very intricate.
On the other hand, in type-IIB ground-states, the exact prepotential is given by
the geometry of
the moduli space of complex structures, which can be calculated using classical
geometry.
Mirror symmetry can be further used \cite{mirror} to solve the analogous
type-IIA problem.

An interesting phenomenon, is that generically, CY manifolds develop some
conifold
(logarithmic) singularities at some submanifolds of their K\"ahler moduli
space.
In a type IIA compactification such singularities appear at tree level and
cannot be smoothed
out by quantum effects since as we have argued there aren't any.
At such conifold points a collection of two-cycles shrinks to zero size.
It was however been pointed out \cite{coni}, that at such points,
non-perturbative states
(D2-branes wrapped around the vanishing cycles) become massless. If we included
them explicitly
in th effective theory, then the singularity disappears.
Alternatively, integrating them out reproduces the conifold singularity.
The message is the type II perturbation theory gives directly the full quantum
effective action
after integrating out all massive degrees of freedom.

In the type-IIB ground-state the conifold singularity appears in the hypermoduli
space.
Here we expect both perturbative and non-perturbative corrections to smooth-out
the singularity.
This has been confirmed in some examples \cite{OV}.

(b) \un{Heterotic perturbation theory}. This type of perturbation theory
applies to ground-states
of heterotic theory compactified on a six-dimensional manifold of SU(2)
holonomy (a prototype is
K3$\times$T$^2$) and type-II asymmetric vacua with (2,0) supersymmetry (ie.
both
supersymmetries
come from the left-movers or right-movers).
In such vacua, the dilaton belongs to a vector multiplet. Thus supersymmetry
implies that
the tree-level geometry of the hypermoduli space is exact.
On the other hand the geometry of the vector moduli space (prepotential)
receives perturbative
corrections only at one-loop, as well as non-perturbative corrections due to
space-time
instantons.
Several such vacua seem to be dual to type-II (1,1) vacua \cite{hetII}.
This duality can be used to determine exactly the geometry both of the vector moduli space
as well as the hypermoduli space.
Moreover, such a map is consistent with the non-renormalization theorems 
mentioned above, and it reproduces the one-loop contribution on the heterotic side
\cite{n=21} as well the Seiberg-Witten geometry in the global limit \cite{n=22}.

One more property should be stressed here: in heterotic-type N=2 vacua, there is no
renormalization of the Einstein term. On the other hand there is a  gravitational
(universal) contribution to the gauge couplings \cite{a1,KK} 
as well as the K\"ahler potential \cite{a2}. This can be thought of as due to 
world-sheet contact terms \cite{a1} or as the gravitational back reaction \cite{KK}.
Its diagrammatic interpretation (for the gauge coupling) is that of a 
space-time contact term where two gauge bosons
couple to the dilaton which couples to a generic loop of particles (see \cite{book} for
more details).

(c)  \un{type-I perturbation theory}. This is the perturbation theory of type II
orientifolds that contain open sectors. Here we have both open and closed unoriented 
Riemann surfaces.
Here the dilaton belongs partly to a vector multiplet and partly to a hypermultiplet.
As a result both the vector moduli space as well as the hypermoduli space receive
corrections perturbatively and non-perturbatively.
Moreover there is another subtlety: there is no universal renormalization
of the gauge couplings and the K\"ahler potential. On the other hand there is 
a one-loop (cylinder) renormalization of the the Einstein term \cite{ABFPT} consistent
with type-I/heterotic duality.

There is a whole series of ``chiral" F-terms $I_g$ that generalize the prepotential and
the two-derivative effective action, \cite{n=2t1,n=2t2,dwit,FI}
\be
I_g=\int d^2\theta ~W^{2g} {\cal F}_{g}(X)
\ee
where
\be
W_{\m\n}^{ij}=F_{\m\n}^{ij}-R_{\m\n\r\s}\theta^i\s_{\r\s}\theta^j+\cdots
\ee
is the supergravity superfield, with the anti-self-dual graviphoton field strength
$F^{ij}_{\m\n}$ and the anti-self-dual Riemann tensor.
The square is defined as $W^2=\epsilon_{ij}\epsilon_{kl}W^{ij}_{\m\n}W_{kl}^{\m\n}$.
The superfields X, stand for the vector multiplet superfields,
$X^I=\phi^I+{1\over 2}F^{I}_{\m\n}\theta^1\s_{\m\n}\theta^2+\cdots$, $X^0$ corresponds to
the graviphoton.

In type II perturbation theory, we can go to the $\s$-model frame by the gauge fixing
condition $X^0=e^{K/2}/g_s$ where $K$ is the K\"ahler potential and $g_s=e^\Phi$ is the
string coupling constant. Then $Z^i=X^i/X^0$ are the true moduli scalars.
From supersymmetry we know that ${\cal F}_g$ must be a homogeneous function of degree $2-2g$.
Thus,
\be
{\cal F}_{g}\sim (X^0)^{2-2g}\tilde {\cal F}_g (Z^i)\sim e^{(1-g)K}(g_s^2)^{g-1}~
\tilde {\cal F}_g(Z^i)
\ee
This implies that such effective terms obtain a contribution only at the g-th order in
type-II perturbation theory.
This was indeed verified by an explicit computation, \cite{n=2t1}.
${\cal F}_0$ is indeed the prepotential that governs the two derivative interactions.
${\cal F}_1$ governs among other things, the $\Rr^2$ terms and obtains contributions from
one loop only.

The (almost) holomorphic threshold is given by the topological partition
function, of a twisted CY $\s$-model \cite{n=2t1}.
The mild non-holomorphicity is due to an anomaly and it provides for recursion relations 
among the various ${\cal F}_g$'s.
They have the form
\be
\pa_{\bar A}{\cal F}_{g}={1\over 2}\bar C_{\bar A\bar B\bar C}e^{2K}G^{B\bar B}
G^{C\bar C}
\left(D_B D_{C}{\cal F}_{g-1}+\sum_{r}D_{B}{\cal F}_r D_{C}{\cal F}_{g-r}\right)
\ee
where $D_A$ is the K\"ahler covariant derivative and $C_{ABC}$ are the holomorphic Yukawa
couplings.
For $g=1$ it is equivalent to
\be
\pa_A\pa_{\bar A}{\cal F}_1={1\over 2}\left(3+h_{11}-{1\over 12}\chi\right)G_{A\bar A}
-{1\over 2}R_{A\bar A}
\ee

Near a conifold point they all become singular in a different fashion as one approaches
the singularity.
Their singularity structure was shown to be captured by the c=1 topological 
matrix model \cite{ghos}.

\subsection{N=4 Supersymmetry\label{n=4}}
\setcounter{equation}{0}

The critical dimension of N=4 supersymmetry is ten. In ten dimensions it
corresponds to a
single Majorana-Weyl supercharge which decomposes to four supercharges upon
toroidal
compactification to four dimensions.
The relevant massless multiplets are the supergravity multiplet (the graviton,
second rank tensor
a scalar, a Majorana-Weyl gravitino and a Majorana Weyl fermion of opposite
chirality
in ten dimensions) and the vector multiplet ( a gauge boson and a Majorana-Weyl
gaugino).
In d dimensions, the supergravity multiplet contains apart from the metric and
second rank
tensor and original scalar, (10-d) vectors (graviphotons).
The vector multiplet contains apart from the vector an extra (10-d) scalars.

The two-derivative effective action is completely fixed by supersymmetry and
the
knowledge of the gauge group.
Its salient features are that it has no scalar potential in ten dimensions
(since there are no
scalars in the theory), and it has a Chern-Simons coupling of the gauge fields
to the second-rank
tensor crucial for anomaly cancellation via the Green-Schwarz mechanism.
The explicit action and a discussion can be found in \cite{book}.

In lower dimensions scalars appear from the components of the metric, second
rank tensor and
gauge fields. There is a potential for the scalars due in particular to the
non-abelian field
strengths. The minima of the potential have flat directions parametrized by
expectation values of
the scalars coming from the supergravity multiplet as well as those coming from
the Cartan
vectors.
These expectation values break generically the non-abelian gauge symmetry to
the maximal
abelian subgroup. At special values of the moduli massive gauge boson can
become massless
and gauge symmetry is enhanced.

Supersymmetry constraints the local geometry of the  scalars in d dimensions to
be that of
the symmetric space O(10-d,N)/O(10-d)$\times$O(N) where N is the number of
abelian
vector multiplets.
Moreover, if we neglect the massive gauge bosons, and focus on the generically
massless fields then the effective action is invariant under a continuous
O(10-d,N) symmetry
under which the metric, second rank tensor and original scalar are inert, while
the vectors
transform in the vector while the scalars in the adjoint.
The O(10-d) part of this symmetry is the R-symmetry that rotates the
supercharges.

The O(10-d,N) symmetry is broken by the massive states.
In string theory a discrete subgroup remains that is generically a subgroup of
O(10-d,N,$\Z$).
The interested reader  can find a more detailed discussion of the above in
\cite{book}

Since supersymmetry and knowledge of the rank of the gauge group completely
fixes the
two-derivative effective action of the massless modes, we expect no
perturbative or
non-perturbative corrections. This has been explicitly verified in various
contexts.
In the context of field theory (global N=4 supersymmetry) it can be shown that
in
the Higgs phase
perturbative corrections vanish, as well as instanton corrections (due to
trivial zero mode
counting) \cite{n=4}. Moreover, in the Higgs phase there are no subtleties with
IR divergences.
In the local case (string theory) perturbative non-renormalization theorems
have been advanced (see \cite{GSW}).

The knowledge of BPS-saturated terms for N=4 supersymmetry is scarce.
The next type of terms  beyond the two derivative ones, are those related by
supersymmetry to $\Rr^2$ (CP-even).
Among these, there are the CP-odd terms  $tr \Rr\wedge \Rr$ (four dimensions) 
and $B\wedge
tr \Rr\wedge \Rr$ (six dimensions).
In a type-II (1,1) setup, there is no tree-level $\Rr^2$ term \cite{GSW,slo}, but there 
is a contribution at one-loop. It has been conjectured \cite{HM} that there are
no further perturbative and non-perturbative corrections. Arguments in favor of this
conjecture were advanced in \cite{6}. In particular this conjecture seems to be in
agreement with heterotic/type II duality.
In the respective heterotic perturbation theory, the $\Rr^2$ term has a tree level
contribution \cite{slo}, but no further perturbative contributions.
One could expect non-perturbative contributions in four dimensions due to Euclidean 
five-brane instantons wrapped around $T^6$ \cite{HM}. This is compatible with the type II
one-loop contribution and heterotic/type II duality.
The situation in type-I perturbation theory seems to be less clear: it is only 
known that there is a non-trivial one-loop (cylinder) contribution to the 
$\Rr^2$ term below ten dimensions \cite{fs}.   
The one-loop correction to the $\Rr^2$ term is proportional to the conformal anomaly.
Moreover the conformal anomaly depends on the ``duality frame" \cite{duff}; 
although in four dimensions a pseudoscalar is dual to a second rank tensor, they
contribute differently to the conformal anomaly; when the scalar contributes 1 the 
second-rank tensor contributes 91.
In the heterotic side we have an antisymmetric tensor and this provides for the vanishing
one-loop result while on the type-II side we are in a dual frame and this gives a non-zero
one-loop result.

There are several other terms with up to eight derivatives that are 
of the BPS saturated
type.
These include $\F^4, \F^2 \Rr^2$ and $\Rr^4 $
 terms \cite{Tseytlin,BaKi,bk2,lec}.
So far, we have been vague concerning the tensor structure of such
terms.
Here, however, we will be more precise \cite{roo,Tseytlin,BaKi}.
There are three types of $\Rr^4$ terms in ten dimensions: $t_8(tr\Rr^2)^2$,
$t_8tr\Rr^4$ and $(t_8t_8-\e_{10}\e_{10}/8)\Rr^4$, where $t_8$ is the
standard eight-index tensor \cite{GSW}
and $\e_{10}$ is the ten-dimensional totally antisymmetric $\e$ symbol.
The precise expressions can be found for example in \cite{Tseytlin}.
There are also the $t_8 tr\Rr^2tr\F^2$, $t_8tr\F^4$ and $t_8(tr\F^2)^2$
terms.
These different structures can be completed in supersymmetric
invariants
\cite{roo,Tseytlin}.
The bosonic parts of these invariants are as follows:
\bs
\be
J_0=\left(t_8t_8+{1\over
8}\e_{10}\e_{10}\right)\Rr^4\;\;\;,\;\;\;I_1=t_8tr\F^4-{1\over
4}\e_{10}Btr\F^4\label{10000}\ee
\be
I_2=t_8(tr\F^2)^2-{1\over 4}\e_{10}
B(tr\F^2)^2
\;\;\;,\;\;\;I_3=t_8tr\Rr^4-{1\over 4}\e_{10}Btr\Rr^4
\label{2}\ee
\be
I_4=t_8(tr\Rr^2)^2-{1\over
4}\e_{10}B(tr\Rr^2)^2\;\;\;,\;\;\;I_5=t_8(tr\Rr^2)(tr\F^2)-{1\over
4}\e_{10}B(tr\Rr^2)(tr\F^2) \pe
\label{3}\ee
\label{inv} \es
As is obvious from the above formulae, apart from the $J_0$
combination, the
other
four-derivative terms are related to the Green--Schwarz anomaly by
supersymmetry.
Thus, in ten dimensions, they are expected to receive corrections only
at one
loop if their perturbative calculation is set up properly (in an
Adler--Bardeen-like scheme).
The $J_0$ invariant is not protected by $N=4$ supersymmetry.
Heterotic/type-II duality in six dimensions implies that it receives
perturbative corrections beyond one loop.
It is however protected in the presence of $N=8$ supersymmetry \cite{kp1}.

The relevant N=4 string vacua are the following:

$\bullet$ Type-I O(32) string theory. It is related by weak-strong coupling duality to the
O(32) heterotic string.

$\bullet$ The heterotic O(32) and E$_8\times$E$_8$ strings.
 
$\bullet$ F-theory on K3. This is an d=8 vacuum and is conjectured to be dual to the
heterotic string compactified on $T^2$. 

$\bullet$ Type IIA on K3. It is conjectured to be dual to the heterotic string on $T^4$.
There are further type-II N=4 vacua in less than six dimensions \cite{6}. 
They have either type II or heterotic duals.

We consider first type II N=4 vacua. The case of F-theory 
compactifications \cite{F} 
stands apart in the sense that it has no conventional perturbation theory.
The $F^4$ couplings were derived recently \cite{FF4} 
using geometric methods that mimic those used
in type-II N=2 vacua.

On the other hand there is no computation so far for such terms in the type-II
string compactified on K3. This is the obvious quantitative test 
of the heterotic/type-II
duality and it is still lacking.

Most of the information is known for heterotic and type-I vacua.
The CP-odd terms in (\ref{inv}) were explicitly evaluated
at arbitrary order of heterotic perturbation theory in \cite{ya}.
There, by carefully computing the surface terms, it was shown
that such contributions vanish for $g>1$.
The CP-even terms are related to the CP-odd ones by supersymmetry
(except
for $J_0$).
If there are no subtleties with supersymmetry at higher loops,
then these terms also satisfy the non-renormalization theorem.
This was in fact conjectured in \cite{ya}.
In view of our previous discussion on the structure of supersymmetry,
we would expect that once supersymmetry is working well at $g\leq 1$,
it continues to work for $g>1$ for a suitable definition of the
higher-genus amplitudes.
It is thus assumed that the CP-even terms do not get
contributions beyond one loop.
On the other hand, the $J_0$ term (which is non-zero at tree level) is
not protected by the anomaly.
Thus, it can appear at various orders in the perturbative expansion.
It can be verified by direct calculation that it does not appear at
one loop on the heterotic side.
However, heterotic/type-IIA duality in six dimensions seems to imply
that there is a two-loop contribution to this term on
the heterotic side.
In all of the subsequent discussion for N=4 ground-states , 
when we refer to $\Rr^4$ terms we
will mean the
anomaly-related tensor structures, $I_3$, $I_4$, which can always be
distinguished from $J_0$.

In the type-I theory several of these terms have been calculated and match what is
expected by heterotic/type-I duality \cite{Tseytlin}.
Most of them appear at tree level (disk) and one loop.
There are subtleties though. The heterotic tree-level $(tr \F^2)^2$ term implies via
duality that there should be a two-loop contribution in the type-I side.
That anomaly related terms obtain two-loop contributions is hardly surprising 
if we recall that supersymmetry (that related CP-even and CP-odd terms)
is not respected by type-I perturbation theory \cite{bk2}.

In the heterotic theory, the potential instanton contributions must come from
configurations that preserve half of the supersymmetry.
Thus the only relevant configuration is the heterotic Euclidean five-brane.
It can provide an instanton provided there is a six- or higher-dimensional compact
manifold to wrap it around.
Thus, there are no non-perturbative contributions to such terms in $d>4$ on the heterotic
side.
In four dimensions we generically expect corrections due to NS5-brane instantons.
There is a prediction of global supersymmetry about $F^4$
thresholds \cite{ds2}: it states that there are no corrections  beyond one loop (in the absence of gravity).
We will give an argument in a subsequent section that the same is implied
in the local (string) case by heterotic/type II duality.

The situation is different on the type-I side.
There the relevant configurations are D1 and D5 branes and provide
instanton corrections already in eight dimensions.

No more BPS saturated terms are known in N=4 ground-states.
It was conjectured in \cite{bk2} that in analogy with lower supersymmetry there is 
an infinite tower of BPS-saturated terms as well in the N=4 case.
This rests on the existence of a tower of topological partition functions
in the topological $\s$-model on K3 \cite{BV} in analogy with the N=2 case.
The leading topological amplitude was shown to correspond to the $\Rr^4$ amplitude 
of the type II string compactified on K3.

\subsection{N=8 supersymmetry\label{n=8}}

N=8 ground-states are toroidal compactifications of the type-II string.
Their critical dimension is eleven, and the master theory is eleven-dimensional
supergravity \cite{cj} expected to describe the strong coupling limit of the type-IIA
theory.
The relevant massless representation in eleven dimensions is the supergravity multiplet
and contains the graviton, a three-index antisymmetric tensor and the gravitino.
  
Like N=4 supergravity all two-derivative effective 
couplings receive no renormalization at all.
Unlike N=4 supergravity, four-derivative and six-derivative effective couplings
do not appear at tree level and do not get renormalized at one-loop. It is expected that 
they are not renormalized at all (even beyond perturbation theory).
There are eight-derivative BPS-saturated terms however. $I_3,I_4$ as in the N=4 case, but 
also $J_0$ in this case. There are also related to the $C\wedge \Rr^4$ coupling of
eleven-dimensional supergravity.
These terms obtain tree level plus one-loop corrections in type-II perturbation theory
\cite{GG2,berk1,pio,grs}.
Beyond perturbation theory they obtain corrections from D-instantons.

\section{Instantons in String Theory\label{insta}}

In field theory with $N\geq 2$ supersymmetry instantons are responsible for all
non-perturbative corrections (in the Coulomb branch at least).
What are instantons in string theory? Despite some prescient papers that touched this
issue \cite{chs} the subject took a definite shape only after the duality revolution.
The central idea is that a string instanton to the zeroth approximation is an 
instanton of the effective supergravity theory.
A very important aspect is that instanton solutions preserve a part of the space-time
supersymmetry.

An important conceptual simplification that occurs in string theory is the direct relation
between space-time instantons and wrapped Euclidean solitonic branes \cite{BBS}.
The concept is rather simple. String theory (or the effective supergravity) contains 
solitonic branes that usually preserve some amount of supersymmetry (BPS branes).
These can be found as classical solutions to the supergravity equations of motion.
They include D-branes, the NS5-brane and the M2 and M5 branes of d=11 supergravity.
For D-branes in particular there is an alternative stringy description as Dirichlet 
branes \cite{Polch}. 
An instanton can be produced by wrapping the Euclidean world-volume of a given brane
around an appropriate compact manifold.

What kind of instanton corrections we expect for BPS-saturated terms was discussed case by
case in the previous section. Here we will stress that depending on the term we will need
instantons  with a given number of zero modes.
However, this analysis needs care.
A typical example is that of multi-instanton configurations. 
In multi-instanton solutions, there are in general more bosonic moduli
describing
relative positions and orientation.
If the multi-instanton leaves some supersymmetry unbroken, there will
be more fermionic zero modes, supersymmetric partners of the bosonic
moduli
related by the unbroken supersymmetry.
 If, however their moduli space contains orbifold singularities, then 
there are contributions localized at the singularities where the number of zero modes is
reduced.
We will see later an explicit example of this in the case of D1-instanton contributions
to four-derivative couplings in type-I string theory.

An important  question to be answered is: What part of the supersymmetry can
an instanton configuration  break?
The answer to this  depends on the particular instanton (Euclidean brane) 
as well as the number of non-compact
dimensions.
It depends crucially on the compact manifold, and the way the Euclidean brane is wrapped
around it \cite{BBS,inbr}.

Can we compare between the instantons we are
using in string theory and standard field-theory instantons?
In field theory, we are usually considering two types of instantons.
The first are instantons with finite action, and a typical example
is the BPST instanton \cite{bpst}, present in non-Abelian
four-dimensional gauge theories.
Examples of the other type are provided by the Euclidean Dirac monopole
in three dimensions, which is relevant, as shown in \cite{poly}, to  the
understanding of the non-perturbative behaviour of three-dimensional
gauge theories in the Coulomb phase.
This type of instanton has an ultra-violet (short-distance)-divergent
action, since it is a singular solution to the Euclidean
equations
of motion.
However, by cutting off this divergence and subsequent renormalization,
it can contribute to non-perturbative effects.
The generalization to the compact gauge theories of higher antisymmetric
tensors was also discussed in the context of (lattice) field theory \cite{peros}.  
Another famous case in the same class is the two-dimensional vortex
of the XY model, responsible for the KT phase transition \cite{KT}.
In four dimensions we also have the BCD merons \cite{mer}, with similar
characteristics, although their role in the non-perturbative
four-dimensional dynamics is not very well understood.

In the context of string theory, we  have these two types of
instantons.
Here, how\-ever, the behaviour seems to be somewhat different.
Let us consider first the heterotic five-brane \cite{chs}.
This solution is intimately connected to BPST instantons in the
transverse
space and is smooth provided the instanton size is non-zero.
At zero size the solution has an exact CFT description but the string
coupling is strong. Non-perturbative effects are important and a
conjecture
has been put forth to explain their nature \cite{zerosize}.
Another type of instanton whose effective field-theory description is
regular
is the D3-brane of type-IIB theory.
On the other hand, the other D-brane instantons have an effective
description that is of the singular type. However, their ultra-violet
divergence
is cured in their
stringy description.
This is already clear in the case of the type-I D1-brane that will be described 
in these lectures, where the effective description is 
singular \cite{dab,typeI1} while the
stringy description
turns out to be regular and in particular, as we will see later,
their classical action is finite.

There seems to be a correspondence of the various field-theory
instantons
to stringy ones. We have already mentioned the example of the heterotic
five-brane, but the list does not stop there.
In \cite{bk} it was shown that the three-dimensional Polyakov QED
instanton
as well as various non-Abelian merons have an exact CFT description
and thus correspond to exact classical solutions of string theory.
Moreover, the three-dimensional instanton can be interpreted as an avatar of
the five-brane zero-size instanton when the theory is compactified to three
dimensions.
Similar remarks apply to the stringy merons, which require the presence
of five-branes with fractional charge \cite{bk}. In that respect
they are solutions of the
singular
type in the effective field theory.
In the context of the string theory, the spectrum of instanton
configurations
is of course richer, since the theory includes gravity.
However, the correspondence of field-theory and some string-theory
instantons
implies that the field-theory non-perturbative phenomena associated
with them, are
already included in a suitable stringy description.

\section{Heterotic/Type-I duality and D-brane instantons}

The conjectured duality \cite{typeI,typeI1,dab,PW}
 between the type-I and heterotic
$Spin(32)/Z_2$ string theories is particularly intriguing. 
The massless spectrum of both theories, in ten  space-time dimensions,
contains the (super)graviton and the  (super)Yang Mills multiplets.
Supersymmetry and anomaly cancellation 
fix  completely  the  low-energy Lagrangian, and more precisely
all two-derivative terms  and  the anomaly-canceling,
four-derivative Green-Schwarz couplings
\cite{GSW,Tseytlin}. One logical possibility, consistent with this
unique low-energy behaviour, could have been that the two theories are
self-dual  at strong coupling.  The conjecture that they
are instead dual to each other implies  that this unique infrared
physics  also has a unique consistent  ultraviolet extrapolation.

    One of the early arguments in favour of this duality \cite{typeI,typeI1,dab}
was that the heterotic string appears as a singular solution of the
type-I theory. Strictly-speaking this is not  an  independent
test of duality. 
Since the two effective actions are related by a field redefinition 
this is not surprising.
The real issue is whether consistency of the
theory forces us to include such excitations in the spectrum.
This can for  instance be argued in the case of  type
 II string theory near a
conifold singularity of the Calabi-Yau moduli space \cite{coni}.

We are not aware of such a direct argument in the case of the heterotic
string solution. What is, however, known is that it
has  an exact conformal description
as a D(irichlet) string of  type-I theory \cite{PW}.
In certain  ways,
D-branes lie between fundamental quanta and smooth solitons
 so, even if we admit  that they  are
intrinsic,  we must still decide on the rules for
including  them  in a semiclassical  calculation.
Do  they contribute, for instance, to loops like
fundamental quanta?  And with what measure and degeneracy should we
weight  their Euclidean trajectories?

Here we will analyse some calculations \cite{BaKi,bk2}
in which these questions can be  answered.
The rules consistent with  duality turn out to be  natural and
simple.   D-strings, like smooth  solitons, 
 do not enter {\it explicitly} in loops
\footnote{A (light) soliton loop can of course
 be a useful approximation
to the exact instanton sum, as is the case  near the strong-coupling
singularities of the Seiberg-Witten solution. For a D-brany discussion see also \cite{neuc}.},
while their  (wrapped)  Euclidean trajectories contribute to
the saddle-point sum, without topological degeneracy if one takes into
account correctly the non-abelian structure of D-branes.

\subsection{Heterotic/Type-I duality in ten dimensions.\label{hetI}}
\setcounter{equation}{0}

We will start our discussion by briefly describing heterotic/type-I duality
in ten dimensions.
It can be shown \cite{sen} that heterotic/type-I duality, along with
T-duality can reproduce all known string dualities.

Consider first the O(32) heterotic string theory.
At tree-level (sphere) and up to two-derivative terms, the (bosonic)
effective
action in the $\s$-model frame is
\be
S^{\rm het}=\int d^{10}x\sqrt{G}e^{-\Phi}\left[
R+(\nabla\Phi)^2-{1\over 12}\hat H^2-{1\over 4}F^2\right]
\,.\label{570}\ee

On the other hand for the O(32) type I string the leading order
two-derivative effective action is
\be
S^{I}=\int d^{10}x\sqrt{G}\left[e^{-\Phi}\left(
R+(\nabla\Phi)^2\right)-{1\over 4}e^{-\Phi/2}F^2-{1\over 12}\hat
H^2\right]
\,.\label{571}\ee
The different dilaton dependence here comes as follows: The Einstein
and dilaton terms come from the closed sector on the sphere ($\chi=2$).
The gauge kinetic terms come from the disk ($\chi=1$). Since the
antisymmetric tensor comes from the RR sector of
the closed superstring it does not have any dilaton
dependence on the sphere.

We will now bring both actions to the Einstein frame,
$G_{\m\n}=e^{\Phi/4}g_{\m\n}$:
\be
S^{\rm het}_E=\int d^{10}x\sqrt{g}\left[
R-{1\over 8}(\nabla\Phi)^2-{1\over 4}e^{-\Phi/4}F^2-{1\over
12}e^{-\Phi/2}\hat H^2\right]
\,,\label{572}\ee
\be
S^{I}_E=\int d^{10}x\sqrt{g}\left[
R-{1\over 8}(\nabla\Phi)^2-{1\over 4}e^{\Phi/4}F^2-{1\over
12}e^{\Phi/2}\hat H^2\right]
\,.\label{573}\ee

We observe that the two actions are related by $\Phi\to -\Phi$ while
keeping the other fields invariant. This suggests that the weak
coupling of one
is the strong coupling of the other and vice versa.
As mentioned earlier the fact that the two effective actions are related by a field redefinition is not surprising. What is 
is interesting though is that the field redefinition here is just an
inversion of the ten-dimensional coupling.
Moreover, the two theories have perturbative expansions that are very
different.

Let us first study the matching of the BPS-saturated high derivative terms
in ten dimensions.
At tree level, the only four-derivative term is the $(tr \F^2)^2$.It is part of the 
Chern-Simons related combination $(tr\F^2-tr \Rr^2)^2$ \cite{slo}.
Via the duality this
term should correspond to a type-I contribution that comes from a genus-3 surface.
This, of course, has never been computed.
At one loop $\F^4$ terms would correspond to disk term in the type-I theory.
Fortunately, the only non-zero one-loop contribution is of the type $tr \F^4$
and agrees with the disk computation.
$(tr \F^2)^2$ is zero at one-loop in the heterotic theory, a good thing since it would be impossible to obtain such a term from the disk (that has a single boundary).
Similar remarks apply to the $\Rr^4$ and mixed terms.

We should stress again here that the matching of the one-loop heterotic terms with 
specific disk and one-loop terms in type-I is not a test of duality.
It is rather a consequence of N=1 supersymmetry and anomaly 
cancellation in ten dimensions.

\vskip 0.2cm
\subsection{One-Loop Heterotic  Thresholds\label{1lhet}}
\setcounter{equation}{0}

 As discussed previously, the terms that will be of interest to us are those obtained
by dimensional reduction from the  ten-dimensional superinvariants, whose 
bosonic parts read \cite{roo,Tseytlin}
\be
\eqalign{
I_1=&t_8tr{\cal F}^4-{1\over4}\e_{10}Btr{\cal F}^4,\quad 
I_2=t_8(tr{\cal F}^2)^2-{1\over 4}\e_{10}B(tr{\cal F}^2)^2\cr
I_3=&t_8tr{\cal R}^4-{1\over 4}\e_{10}Btr{\cal R}^4 ,\quad
I_4=t_8(tr{\cal R}^2)^2-{1\over4}\e_{10}B(tr{\cal R}^2)^2\cr
I_5=&t_8(tr{\cal R}^2)(tr{\cal F}^2)-
{1\over4}\e_{10}B(tr{\cal R}^2)(tr{\cal F}^2)\pe\cr
}\ 
\label{1}
\ee
These are special because they contain anomaly-canceling
CP-odd  pieces. As a result  anomaly cancellation fixes entirely their
coefficients in both the heterotic and the type I
 effective actions in ten dimensions. 
Comparing these coefficients
is not therefore  a test of duality, but rather of the fact that both
these theories are consistent \cite{Tseytlin}.
In lower dimensions things are different: the coefficients of the various
terms,  obtained from a single ten-dimensional superinvariant through dimensional
reduction,  depend on the compactification  moduli. 
Supersymmetry is expected to
relate  these coefficients  to each other, but is not   powerful
enough so as to fix them completely.
This is analogous
to the case of  N=1 super Yang-Mills in six dimensions:
the two-derivative
gauge-field  action is uniquely fixed, but after toroidal compactification
to four dimensions, it  depends on a holomorphic prepotential which
supersymmetry alone cannot determine.

 On the heterotic side there are good reasons to believe that these
dimensionally-reduced 
 terms receive only one-loop corrections. To start with, this is
true  for their  CP-odd anomaly-canceling pieces 
\cite{ya}. 
Furthermore it has been
argued in the past \cite{cato}
that there exists a  prescription for treating  supermoduli, 
which  ensures that  space-time supersymmetry commutes with
the heterotic genus expansion, at least for vacua with more than
four  conserved supercharges\footnote{A  notable exception 
are compactifications with a naively-anomalous U(1) factor
\cite{dterm,2loop}.}.
Thus, we may plausibly assume that
there are no higher-loop corrections to the terms of interest.
Furthermore, the only identifiable supersymmetric instantons
are the heterotic five-branes. These do not contribute in
  $d>4$ uncompactified dimensions, since they have  no finite-volume
6-cycle to  wrap around. Non-supersymmetric instantons, if they exist,
have  on the other hand  too many fermionic zero modes to
make  a non-zero contribution.  It should be noted that these
arguments do not apply to  the
 sixth superinvariant \cite{roo,Tseytlin}
\be
J_0= t_8 t_8 {\cal R}^4+{1\over 8}\e_{10}\e_{10}
{\cal R}^4 ,
\ee
which is not related to the anomaly.
 This receives as we will mention
below both perturbative and non-perturbative corrections.

The general form of the heterotic one-loop corrections to these
couplings is
\cite{Schellekens,Lerche}
\be
{\cal I}^{het}  =
-{\cal N}\
  \int_{ { F}}{d^2\tau \over
\tau_2^2}\; (2\pi^2 \tau_2)^{d/2} \   \Gamma_{d,d}\
{\cal A}({\cal F},{\cal R}, \tau)
\label{5}\ee
where ${\cal A}$ is an  (almost) holomorphic modular form
of weight zero
related to the elliptic genus, ${\cal F}$ and ${\cal R}$
 stand for the gauge-field
strength and curvature two-forms, $\Gamma_{d,d}$ is the lattice sum
over  momentum and winding modes
for  $d$ toroidally-compactified dimensions, ${ F}$ is the usual
fundamental domain,
 and 
\be
{\cal N} = {V^{(10-d)}\over 2^{10} \pi^6}
\ee
 is a
normalization that includes the volume of the uncompactified
dimensions \cite{BaKi}.
To keep things
 simple we have taken vanishing  Wilson lines on the
$d$-hypertorus, so that the sum over momenta ($p$) and windings ($w$),
\be
\Gamma_{d,d}  =
  \sum_{p,w}
 e^{-{\pi\tau_2\over 2} ( p^2 + w^2/\pi^2)
+ i \tau_1 p\cdot w} \ \ ,
\ee
factorizes inside the integrand.  Our conventions are
\be
 \alpha' = \frac{1}{2} \ ,\ \  q= e^{2\pi i\tau}\ ,
\ \  d^2\tau = d\tau_1 d\tau_2 \
\ee
while   winding  and   momentum  are normalized so that
$p\in {1\over L} {\Bbb Z}$ and
$w\in 2\pi L\ {\Bbb Z}$
for a circle of radius $L$.  The
Lagrangian form of the above lattice sum,
obtained by a Poisson resummation, reads
\be
\Gamma_{d,d}
 = \Bigl({2\over \tau_2}\Bigr)^{d/2}
\sqrt{\det G} \sum_{n_i,m_i \in {\Bbb Z}}
e^{-{2\pi\over\tau_2}\sum_{i,j}
 (G+B)_{ij} ( m_i \tau-n_i) ( m_j \bar\tau-n_j)}
\ee
with $G_{ij}$ the metric and $B_{ij}$ the (constant)
antisymmetric-tensor background
on the compactification torus. For a circle of radius $L$ the metric
is $G= L^2$.

The  modular function   ${\cal A}$ inside the integrand
  depends on the
 vacuum. It is {\it quartic, quadratic} or {\it linear} in ${\cal F}$ and
 ${\cal R}$,
for vacua with {\it maximal, half} or a {\it quarter}
 of unbroken supersymmetries.
The corresponding amplitudes have the property of saturating
exactly the fermionic zero modes in a Green-Schwarz light-cone
formalism, so that the contribution from left-moving oscillators
cancels out \cite{Lerche}\footnote{Modulo the regularization, ${\cal A}$
is in fact the appropriate term in the weak-field expansion of the
elliptic genus \cite{Schellekens,Lerche,Windey,ellwit}}.
 In the covariant NSR formulation this same
fact follows from $\vartheta$-function
 identities. 
\setcounter{footnote}{0}
As a result ${\cal A}$ should
have been
holomorphic in   $q$, but the use of a   modular-invariant
regulator  introduces some
extra  $\tau_2$-dependence \cite{Lerche}.
As a result ${\cal A}$ takes the  generic form 
of a finite polynomial in $1/\tau_2$,  with coefficients that have
Laurent expansions with at most simple poles in $q$,
\be
{\cal A}({\cal F},{\cal R},\tau) =  \sum_{r=0}^{r_{max}}\
 \sum_{n=-1}^\infty  {1\over\tau_2^r}\; q^n \
{\cal
A}^{(r)}_n({\cal F},{\cal R})
. \label{exp}
\ee
The poles in $q$ come from the would-be tachyon. Since this is not
charged under the gauge group, the poles are only present in the purely
gravitational terms of the effective action. This can be verified
explicitly in eq. (\ref{genus}) below.
The $1/\tau_2^r$ terms  play an important role in what follows.
They come from corners of the moduli space where vertex operators,
whose fusion can produce a massless state, collide. Each pair of
colliding operators  contributes  one factor of $1/\tau_2$.
For maximally-supersymmetric vacua the effective action of interest
starts with terms having four external legs, so that $r_{max} = 2$.
For vacua respecting  half the supersymmetries (N=1 in six dimensions
or N=2 in four) the one-loop effective action starts with terms
having two external legs and thus $r_{max} = 1$.

Much of what we will say in the sequel depends only on the above
generic properties of ${\cal A}$.  It will apply  in particular
in  the most-often-studied case of  four-dimensional
vacua with N=2.  For definiteness we will, however, focus
our attention to
the toroidally-compactified SO(32) theory, for which
 \cite{Schellekens,Lerche}
\be
\eqalign{
\ \ \  \cA( {\cal F},{\cal R},\tau )= &\  t_8\;  tr{\cal F}^4
\;+\;\frac{1}{2^7\cdot 3^2\cdot 5} \   {E_4^3\over \eta^{24}}\  t_8\;
tr{\cal R}^4
\;+\; {1\over 2^9\cdot 3^2} {\hat E^2_2 E_4^2\over \eta^{24}}
t_8\; ( tr{\cal R}^2)^2
 \cr &
+{1\over 2^9\cdot 3^2}\; \Bigl[  {E_4^3\over \eta^{24}} +
 {\hat E^2_2 E_4^2\over \eta^{24}}
-2 {\hat E_2E_4E_6\over \eta^{24}} -2^7\cdot
3^2\Bigr]\   t_8\; ( tr {\cal F}^2)^2 \cr \ \ \ \ \ &
+ {1\over 2^8\cdot 3^2}\; \Bigl[   {\hat E_2E_4E_6\over \eta^{24}}
 - {\hat E^2_2 E_4^2\over \eta^{24} }
\Bigr]\   t_8\;   tr {\cal F}^2   tr {\cal R}^2  \pe \cr} \
\label{genus}
\ee
Here $t_8$ is the well-known tensor appearing in four-point amplitudes
of the heterotic string \cite{GSW},
and  $E_{2k}$ are  the
Eisenstein series which are (holomorphic for $k>1$)
 modular forms of weight $2k$. Their explicit expressions are
 collected for convenience in the appendices of \cite{ko}.
The  second Eisenstein series $\hat E_2$ is special, in that it
 requires a non-holomorphic regularization.
 The entire non-holomorphicity of ${\cal A}$ in  eq. (\ref{genus}),
arises through  this modified Eisenstein series.

In the toroidally-compactified  heterotic string
all one-loop  amplitudes with fewer
than four external legs vanish identically \cite{AS}.
Consequently  eq. (\ref{5}) gives directly the effective action,
without
the need to subtract  one-particle-reducible diagrams,  as is the case
at tree level \cite{slo}. Notice also that this four-derivative
effective action has infrared divergences  when more than one
dimensions are
compactified.  Such IR divergences can be regularized in
a modular-invariant way with
a curved background \cite{KK,chem}. This should be kept in mind,
 even though for the sake of simplicity we will be working
in this paper  with
unregularized expressions.

\vskip 0.8cm

\subsection{One-loop Type-I Thresholds\label{1lI}}
\vskip 0.2cm
\setcounter{equation}{0}

%%%%%%%%%%%%%%%%%%%%%%%%%%%%%%%%%%%%%%%%%%%%%%%%%%%%%%%%%%%%%%%%%%%%
% Type I   new section

 The one-loop type-I effective action has the form
\be
{\cal I}^{I}  = -{i\over 2}({\cal T} + {\cal K} + {\cal A} + {\cal M})
\ee
corresponding to the contributions of the torus, Klein bottle,
annulus and M{\"o}bius strip diagrams. Only the last two surfaces
(with boundaries) contribute to the ${\cal F}^4$, $({\cal F}^2)^2$
and ${\cal F}^2{\cal R}^2$ 
terms of the action. The remaining two pure gravitational
terms may also receive  contributions from the torus and from the
Klein bottle. Contrary to what happens on the heterotic side,
this one-loop calculation is corrected by  both
higher-order perturbative and non-perturbative contributions.

  For the sake of completeness we review here the calculation of pure
gauge terms following refs. \cite{BF,BaKi}. To the order of
interest only the short BPS multiplets of the open string spectrum
contribute. This follows from the fact that the wave operator
in the presence of a background magnetic field ${\cal F}_{12}=
{\cal B}$  reads
\be
{\cal O} = M^2+ (p_{\perp})^2 + (2n+1)\epsilon + 2\lambda \epsilon
\ee
where $\epsilon \simeq {\cal B} + o({\cal B}^3)$ is a non-linear
function of the field, $\lambda$ is the spin operator projected
onto the plane (12),  $p_{\perp}$
denotes the momenta in the  directions $034\cdots  9$,
 $M$ is a string mass
and $n$ labels the Landau levels. The one-loop free energy thus formally
reads
\be
{\cal I}^{I} = - {1\over 2}
 \int_0^\infty
{dt\over t}\ 
Str\   e^{-{\pi
t\over 2}{\cal O}} 
\ee
where the supertrace stands for a sum over all bosonic minus
fermionic states of the open string, including a sum over 
the Chan-Paton charges, the center of mass
positions and momenta, as well as over the Landau levels.

Let us concentrate on the spin-dependent term 
inside the integrand, which can be expanded for weak field
\be
e^{-\pi t  \lambda\epsilon }= \sum_{n=0}^\infty
 {\left(-\pi t \right)^n\over n!}\left(\lambda\epsilon\right)^n \pe
\ee
The $n<4$ terms vanish for every supermultiplet because of the
properties of the helicity supertrace \cite{BaKi},
while  to the $n=4$ term  only short BPS
multiplets can contribute. The only short multiplets in  the perturbative
spectrum of the  toroidally-compactified
 open string are the SO(32)  gauge bosons and their
Kaluza-Klein dependents. It follows after some straightforward algebra
 that the special ${\cal F}^4$ terms of interest are given by
the following (formal) one-loop super Yang-Mills expression 
\be
{\cal I}^{I} = -{V^{(10-d)}\over
 3\cdot 2^{12}\pi^4}\  
 \int_0^\infty {dt\over t} (2\pi^2 t)^{{d\over 2}-1}
\sum_{p\in {}^*\Gamma}  e^{-\pi t p^2/2}\  \times\  t_8 {\rm Tr}_{adj}
 {\cal F}^4
\ee
where ${}^*\Gamma$ is the lattice of Kaluza-Klein momenta on
a $d$-dimensional torus, and the trace is in the adjoint representation
of SO(32).

 This expression is quadratically UV  divergent, but
in the full string theory one must remember to (a) regularize contributions
from the annulus and M{\"o}bius uniformly in the {\it transverse}
closed-string  channel, and (b) to subtract the one-particle-reducible
diagram corresponding to the exchange of a massless (super)graviton
between two $tr{\cal F}^2$ tadpoles, with the trace being here
in the fundamental representation of the group.
 The net result can be summarized easily, after a Poisson resummation
from the open-channel Kaluza-Klein momenta to the closed-channel
windings, and  amounts to simply subtracting the contribution of the
zero-winding sector 
\cite{BF,BaKi}. 
 Using also the fact
that ${\rm Tr}_{adj}
 {\cal F}^4 = 24 tr {\cal F}^4 + 3 (tr {\cal F}^2)^2$ we thus derive
the final one-loop expression on the type-I side
\be
{\cal I}^{I} = -{V^{(10)}\over
  2^{10}\pi^6}\  
 \int_0^\infty {dt\over t^2} 
\sum_{w\in \Gamma\backslash\{0\}} 
 e^{-  w^2/2\pi t}\  \times\  t_8 \Bigl( tr {\cal F}^4 + {1\over 8}
 (tr {\cal F}^2)^2 \Bigr)
 \ .
\ee
The conventions for momentum and winding are the same as in
the heterotic calculation of the previous section.

  The calculation of the gravitational terms is more involved
because we have no simple background-field method at our disposal.
It can be done in principle
 following the method described in ref. \cite{ABFPT}. There is one
particular point we want to stress here: 
if the  one-loop heterotic calculation is exact, and assuming that
duality is valid, there should be no world-sheet instanton corrections
on the type-I side. Such corrections would indeed translate to
non-perturbative contributions  in the heterotic string
\cite{silver}, and we have
just argued above that there should not  be any. The dangerous diagram
 is the torus which can wrap non-trivially around the compactification
manifold.
The type-I torus diagram  is on the other hand
identical to the type IIB one,  assuming there are  only graviton insertions.
This latter diagram was  explicitly calculated  
in eight uncompactified dimensions  in ref. \cite{kp1},  confirming 
our  expectations: the CP-odd invariants
  only  depend  on the complex
structure of the compactification torus, but not  on
its K{\"a}hler structure. This is not true for the CP-even invariant
$J_0$.

\vskip 0.8cm

\subsection{Circle  Compactification\label{d9}}
\vskip 0.2cm
\setcounter{equation}{0}

Let us  begin now our comparison  of the effective actions  with the
simplest situation, namely
  compactification on a circle. There are no world-sheet or D-string
instanton contributions 
in this case, since Euclidean world-sheets have no
finite-area
manifold in target space
to wrap around.  Thus, the one-loop heterotic
amplitude  should be expected to match
with a perturbative calculation on the type-I side.
This sounds at first puzzling,  since the heterotic theory  contains
infinitely more charged BPS multiplets than the type-I theory in its
perturbative spectrum. Indeed, one can combine
any state of the $SO(32)$ current
algebra  with appropriate $S^1$-winding and momentum,
so as to satisfy   the level-matching condition of physical states.
The heterotic theory thus contains short multiplets
in arbitrary representations of the gauge group.

The puzzle is resolved by  a well-known trick, used 
previously in the study of string thermodynamics \cite{McC,KS}, and which  
 trades  the winding sum for an unfolding of the
fundamental domain  into the half-strip,
$-{1\over 2} < \tau_1 < {1\over 2}$ and $\tau_2>0$.
The trick works  as follows: starting with
the Lagrangian form of the heterotic lattice sum,
 \be
(2\pi^2\tau_2)^{1/2}\  \Gamma_{1,1}  = 2\pi L
  \sum_{(m,n) \in {\Bbb{Z}}^2 }
 e^{- 2\pi L^2\vert m\tau -n\vert^2/\tau_2}
\pe \ee
one decomposes  any non-zero pair of  integers   as
$(m,n) = (jc,-jd)$, where $j$ is their greatest common divisor
(up to a sign). We will denote the set of all relative primes
$(c,d)$, modulo an overall sign, by 
${\cal S}$. 
The lattice sum can thus  be written as
\be
({2\pi^2\tau_2})^{1/2} \  \Gamma_{1,1}
 = 2\pi L \Bigl[\  1+
 \sum_{j\in  {\Bbb{Z}}\backslash \{0\}}\   \sum_{(c,d)\in {\cal S} }
 e^{- 2\pi L^2 j^2\vert c\tau +d \vert^2/\tau_2} \Bigr] \pe
\ee

Now the set ${\cal S}$ is in one-to-one correspondence with
all modular
transformations,
\be
{\tilde \tau} = {a\tau +b\over c\tau+d}\ \  \Longrightarrow\
{\tilde\tau}_2 = {\tau_2\over \vert c\tau+d\vert^2} \
\ee
such that
$
-{1\over 2} < {\tilde\tau}_1 \le {1\over 2}\ $.
 Indeed  the condition  $ad-bc=1$ has a solution  only if
 $(c,d)$ belongs  to  ${\cal S}$,
and the solution is unique modulo a shift and an irrelevant sign
\be
\left( \matrix{a  & b \cr c & d}\right)
 \rightarrow \pm  \left( \matrix{1 &{ l} \cr 0 & 1}\right)
\left( \matrix{a  & b \cr c & d}\right)\ .
\label{transfo}
\ee
By choosing  ${ l}$  appropriately we may always bring $\tilde\tau$
inside the strip, which establishes the above  claim.

Using the modular invariance of ${\cal A}$, we can thus suppress the
sum
over $(c,d)\in \cal S$ and unfold  the integration regime
for the $j\not= 0$ part of the expression.  This gives
\be
{\cal I}^{het} =
 -{V^{(9)} L\over 2^9 \pi^5}\; 
\Biggl[  \int_{{ F}}{d^2\tau \over
\tau_2^2}\;{\cal A}
\ + \
\int_{\rm strip} {d^2\tau\over\tau_2^{\ 2}}\;
 \sum_{j \neq 0}
 e^{- 2\pi L^2
j^2/\tau_2 }\;{\cal A}\; \Biggr]
   \pe
\label{trick}
\ee
There is  one subtle point in this  derivation
\cite{KS}:
convergence of the original threshold integral, 
 when ${\cal A}$ has a
 ${1\over q}$ pole\footnote{(Physical) massless states do not lead to
IR divergences in four-derivative  operators in nine dimensions},
 requires that we integrate 
 $\tau_1$ first in the  $\tau_2\to\infty$ region.
Since constant $\tau_2$ lines
transform  however non-trivially under SL(2,{\bf Z}),
the integration over the entire strip would have to be supplemented
by a highly  singular  prescription.
The problem could be  avoided
 if  integration of the $m\not= 0$ terms in the Lagrangian
 sum (i.e. those  terms that required a change of integration variable)
 were  absolutely convergent. This is the case for $L>1$,
so 
 expression (\ref{trick}) should only be trusted in
this  region.

\begin{figure}
\begin{center}
\leavevmode
\epsfxsize=16cm

\epsfbox{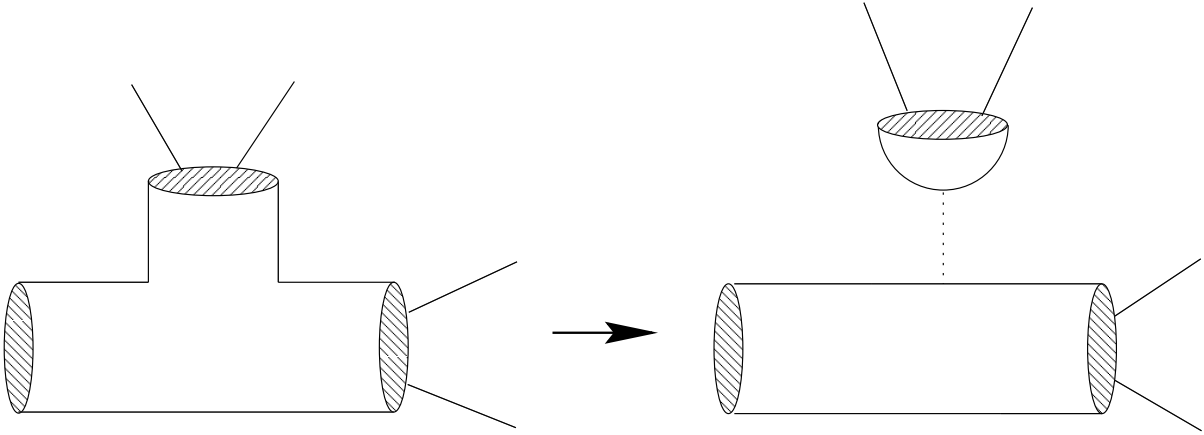}
\end{center}
\caption[]{
A type-I diagram with Euler characteristic
$\chi=-1$. This contributes to the
  $(tr{\cal F}^2)^2$ piece of the
effective action,  only in degeneration limits such as the one depicted above.}
\label{f1}\end{figure}

 Let us now proceed to evaluate this  expression. The fundamental
domain
integrals can be performed explicitly by using the formula
\cite{Lerche}
\be
 \int_{ F} {d^2\tau\over \tau_2^2}
 ({\hat E}_2)^r  \Phi_{r} =
\  {\pi\over 3(r+1)} [c_0 -24(r+1) c_{-1}]
\ee
where
\be
\Phi_{r}(q) =  \sum_{n=-1}^\infty c_n q^n
\ee
is any modular form of weight $-2r$ which is holomorphic everywhere
except possibly for a simple pole at zero.
As for the strip integration, it picks up only the ${\cal O}(q^0)$
term in the expansion of ${\cal A}$.
Modulo  the non-holomorphic regularization, 
only the SO(32)  gauge bosons contribute to the elliptic
genus at this order, in agreement precisely with the result of  the
type-I side!
 For $k\ge 1$ let us  define more generally
\be
 \; \int_0^\infty {d\tau_2\over \tau_2^{1+k}} \sum_{j\not= 0}
 e^{-2\pi L^2 j^2/\tau_2} =  {2 \Gamma(k) \zeta(2k)
 \over (2\pi L^2)^k}\ \equiv {N_k\over L^{2k}},
\label{Zag}
\ee
where $L$ is the radius of the compactification circle.
The one-loop SO(32) heterotic action takes finally the form
\be
\eqalign{
 {\cal I}^{het} =  -{ V^{(10)} \over 2^{10}\pi^6}&\;
\Biggl\{ \;\; 
  {\pi\over 3}  \Bigl[
  \; {\cal F}^4\; - {1\over 8}\; {\cal F}^2 {\cal R}^2
  + {1\over 8}\; {\cal R}^4 \;
+\; {1\over 32} \left(  {\cal R}^2  \right)^2  \Bigr] + \cr
+  & {N_1\over L^2} 
 \Bigl[
   \; {\cal F}^4\; + {1\over 8}  ( {\cal F}^2)^2\;
-{5\over 16} \;  {\cal F}^2 {\cal R}^2  +  {31\over
240}  {\cal R}^4 \;
+\frac{19}{ 192 }\;  \left(  {\cal R}^2  \right)^2  \Bigr] - \cr
-{5\over 16\pi}\times  &
 {N_2\over L^4} 
 \Bigl[
3  ( {\cal F}^2)^2\;
 - {5} \;  {\cal F}^2  {\cal R}^2  +\; 2
  \left(  {\cal R}^2  \right)^2  \Bigr]
 \ \;  +\;\  {21 \over 64\pi^2}\times  
 {N_3\over L^6}
 ( {\cal F}^2  -  {\cal R}^2)^2\;  \Biggr\} \pe}
\label{s1}
\ee
To simplify notation we have written here ${\cal F}^4$ instead of
 $t_8\; tr {\cal F}^4$,
  $({\cal F}^2)^2$ instead of  $t_8\;  tr {\cal F}^2 tr {\cal F}^2$
etc.

We have
 expressed the result as an expansion in inverse
powers of the compactification volume.
Since the heterotic/type-I duality map
transforms ($\sigma$-model) length scales as
\be
L^2_h = L^2_I/\lambda_I
\ee
with $\lambda_I$ the open-string loop counting parameter, this
expansion
can be translated to a genus expansion on the type-I side.
The Euler number
 of an non-orientable surface is given by
$\chi=2-2g-B-C$
where $g$ is the number of holes, $B$ the number of boundaries
and $C$ the number of cross-caps.
The leading term corresponds to the disk and projective plane 
diagrams
and is completely fixed
by ten-dimensional
supersymmetry and  anomaly cancellation \cite{Tseytlin}.
To check this, one must remember to transform
the metric in  both $V^{(10)}$ and the tensor $t_8$ appropriately.
Notice that the type-I sphere diagram, which is the same as in type IIB,
only contributes to the $J_0$ invariant which we are not considering
here.
The subleading o($L^{-2}$) terms correspond to the annulus, M{\"o}bius
strip, Klein bottle and torus diagrams, all with $\chi = 0$. 
For zero background
curvature these agree  with the type-I calculation \cite{BaKi}
as described  in section \ref{1lI}.

  The last two  terms in the expansion (\ref{s1}) correspond 
to diagrams with $\chi = -1,-2$. These contributions must be there
if the  duality map of ref. \cite{typeI} does not receive higher-order
corrections. Such corrections could anyway always be absorbed
by redefining fields on the type-I side, so that if duality holds,
there must exist  some regularization scheme in which these higher-genus
contributions do arise. These terms do on the other hand come from
the boundary of moduli space. For instance the $\chi =-1$ contribution
to the $({\cal F}^2)^2$ term comes from the boundary of moduli
space shown in figure 1. It could thus be conceivably 
eliminated in favour of some lower-dimension operators in the effective
action.  
 
    It is in any case striking that a single heterotic diagram contains
contributions from different topologies on the type-I side. 
Notice in particular that the divergent $w=0$ term in the one-loop
field theoretic calculation,  regularized on the heterotic side
by replacing  the strip by a fundamental domain, is regularized on
the type-I side by replacing the annulus by the disk.

%%%%%%%%%%%%%%%%%%%%%%%%%%%%%%%%%%%%%%%%%%%%%%%%%%%%%%%%%%%%%%%%%%%%%%

\subsection{ Two-torus Compactification\label{d=8}}

\vskip 0.2cm

 The next simplest situation corresponds to compactification on a
two-dimensional torus. There are in this case world-sheet
instanton contributions  on the heterotic side, and our aim in
this and the following sections will be to understand them as
(Euclidean)
D-string trajectory  contributions on the type-I side.
 The discussion can
be extended with little effort to toroidal compactifications in
lower than eight dimensions. New effects are only expected to arise
in  four or fewer  uncompactified dimensions, where the solitonic
heterotic instantons  or,  equivalently, 
 the type-I  D5-branes  can   contribute.

 The target-space torus 
 is  characterized by  two complex moduli,
the  K{\"a}hler-class
\be
 T = T_1 + iT_2 = \frac{1}{\alpha' } (B_{89} + i\sqrt{G})
\ee
and the complex structure
\be
U = U_1 + i U_2 = ( G_{89} + i\sqrt{G} )/ G_{88} \ ,
\ee
where $G_{\mu\nu}$ and $B_{\mu\nu}$ are the
 $\sigma$-model metric and antisymmetric
tensor  on the heterotic side.
The one-loop  thresholds now read
\be
{\cal I}^{ het}=
{V^{(8)}\over 2^9 \pi^4} \; \int_{ F}{d^2\tau\over
  \tau_2}\Gamma_{2,2}\;  {\cal A}({\cal F},{\cal R},\tau) \ , 
\label{dd8}\ee
where the lattice sum takes the form  \cite{DKL}
\be
    {\Gamma_{2,2}}  =
{ T_2\over \tau_2 }\times
 \sum_{ M \in {\rm Mat}(2 \times 2, {\Bbb{Z}}) }
 e^{ 2\pi i  T {\rm det}M }
e^{- \frac{\pi T_2 }{ \tau_2 U_2 }
\big| (1\; U)M  \big( {\tau \atop -1} \big) \big| ^2 } \pe
\label{DKL}
\ee
The exponent  in the above sum is (minus) the Polyakov action,
\be
S_{\rm Polyakov} = {1\over 4\pi\alpha^\prime} \int d^2 \sigma ( \sqrt{g}
  G_{\mu\nu}g^{\alpha\beta}\partial_\alpha X^\mu \partial_\beta X^\nu
+i  B_{\mu\nu}\epsilon^{\alpha\beta}
\partial_\alpha X^\mu \partial_\beta X^\nu )\ ,
\ee
evaluated 
for the topologically non-trivial mapping of the string world-sheet
onto the target-space torus,
\be
\left(\matrix{X^8\cr X^9\cr}\right)=
 M \left(\matrix{\s^1\cr \s^2\cr}\right)\equiv
\left(\matrix{m_1&n_1\cr
m_2&n_2\cr}\right)\left(\matrix{\s^1\cr \s^2\cr}\right)
 \  \pe
\label{wrap}
\ee
The entries of the matrix $M$ are  integers, and both target-space
and world-sheet coordinates take values in the (periodic) interval
$(0, 2\pi]$. To verify the above assertion one needs to use
the metrics
\be
 G_{\mu\nu} = {\alpha^\prime T_2\over U_2}\; 
\left(\matrix{ 1 & U_1 \cr U_1 & \vert U\vert^2 \cr}
\right)  \; ,
\ \ \ 
g^{\alpha\beta} = {1\over \tau_2^{\ 2}
}\;  \left(\matrix{ \vert\tau\vert^2 & -\tau_1 \cr
-\tau_1 & 1 \cr}\right) \ . 
\ee

The Polyakov action is invariant under global reparametrizations
of the world-sheet,
\be
\left( \matrix{\sigma^1 \cr \sigma^2 \cr}\right)
\rightarrow
\left( \matrix{a & -b\cr -c & d\cr}\right)
\left( \matrix{\sigma^1 \cr \sigma^2 \cr}\right)\ ,
\ee
which transform
\be
\tau \rightarrow    {a\tau+b\over c\tau +d}\ , \ \ {\rm and}\ \ 
M \rightarrow  
 M \ \left( \matrix{d & b\cr c & a\cr}\right) \ 
\pe
\ee
Following Dixon, Kaplunovsky and Louis \cite{DKL}, we
decompose the set of all matrices $M$  into orbits of PSL(2,{\bf Z}),
which is the group of the above transformations up to an overall sign.
 There are three types of orbits,
$$
\eqalign{& {\rm invariant}:  \ M=0 \cr
&  {\rm degenerate}: \  {\rm det} M = 0 ,\;  M\not= 0   \cr
&  {\rm non-degenerate}: \  {\rm det} M \not= 0  \cr}
$$
A  canonical choice of representatives for the  degenerate orbits
is
\be
M  = \left( \matrix{0  & j_1\cr 0  & j_2 \cr}\right)
\ee
where the
 integers $j_1,j_2$ should not both vanish,
but are  otherwise arbitrary.
Distinct elements of  a degenerate orbit are in one-to-one correspondence
with the set ${\cal S}$, i.e.
with modular transformations that map the fundamental domain inside
the strip, as in section \ref{d9}.
In what concerns the  non-degenerate
orbits, a canonical choice of representatives is
\be
 M = \pm \left(\matrix{ k& j\cr 0&p\cr}\right) \ \ 
{\rm with}\ \   0\le j <k \quad, \quad \ p\not= 0\ .
\label{nondeg}
\ee 
Distinct elements of a non-degenerate orbit
 are  in one-to-one correspondence with the fundamental
domains of $\tau$ in the upper-half  complex plane.

Trading  the sum
over orbit elements for  an extension of the integration region
of $\tau$, we can thus express eqs. (\ref{dd8},\ref{DKL}) as
follows  

\be
\eqalign{
{\cal I}^{het} =
-{V^{(8)}  T_2 \over 2^9 \pi^4}  \times &
\Biggl\{  \int_{F}{d^2\tau \over
\tau_2^2}  {\cal A} 
\ + \
\int_{\rm strip} {d^2\tau\over\tau_2^{\ 2}}
\sum_{(j_1,j_2) \neq (0,0)}
 e^{- \frac{\pi T_2 }{ \tau_2 U_2 }
\big| j_1+j_2U \big| ^2 }
\;{\cal A} \cr
&+
2\; \int_{\Bbb C^+} {d^2\tau\over\tau_2^{\ 2}}
 \sum_{{0 \leq j<k} \atop { p\neq 0}}
        e^{2\pi i Tpk}
        \; e^{- \frac{\pi T_2 }{ \tau_2 U_2 }
        \big|k\tau - j-pU \big| ^2 }\; {\cal A}
\Biggr\} \ \equiv {\cal I}_{pert} +{\cal I}_{inst} .
\cr}
\label{3terms}
\ee
The three terms inside the curly brackets are constant,
power-\discretionary{}{}{}sup\-pres\-sed
and exponen\-tially-\discretionary{}{}{}suppressed in the large
compactification-volume 
limit. They correspond to tree-level, higher perturbative and
 non-perturbative, respectively, contributions
on the type-I side. The discussion of the perturbative contributions
follows exactly the analogous discussion in section \ref{d9}. The only
difference is the replacement of eq. (\ref{Zag}) by
\be
\eqalign{
\int_0^\infty {d\tau_2\over\tau_2^{\ 1+k}}\;\sum_{(j_1,j_2) \neq (0,0)}
 e^{- \frac{\pi T_2 }{ \tau_2 U_2 }
\big| j_1+j_2U \big| ^2 }
=&
\Gamma(k) \left({U_2\over\pi T_2}\right)^k \; 
\sum_{(j_1,j_2) \neq (0,0)} \vert j_1 + j_2 U\vert^{-2k}\cr
=& {2 \Gamma(k) \zeta(2k)\over (\pi T_2)^k} E(U,k) . \cr}
\ee
where $E(U,k)$ are generalized Eisenstein
 series.
In the open-string channel of the type-I side this takes into
account properly the (double) sum over Kaluza-Klein momenta 
\cite{BaKi}. Notice that the holomorphic anomalies  in ${\cal A}$ 
lead again to higher powers of the inverse volume, which translate
to higher-genus contributions on the type-I side. Notice also that
the $k=1$ term has a logarithmic infrared divergence, which must be
regularized appropriately, as discussed in the introduction.

 We turn now to the  novel feature of eight dimensions, namely
the contributions  of world-sheet instantons. 
Plugging in the expansion (\ref{exp}) of the elliptic genus, we
are lead to consider the integrals
\be
I_{n,r} = 
\int_{\Bbb C^+} {d^2\tau\over\tau_2^{\ 2}}
         \; e^{- \frac{\pi T_2 }{ \tau_2 U_2 }
        \big|k\tau - j-pU \big| ^2 }\; {1\over \tau_2^{\ r}} e^{2i\pi\tau n}
\label{form2}
\ee
Doing first the (Gaussian) $\tau_1$ integral, 
one finds after some rearrangements
\be
I_{n,r} = {1\over k} \sqrt{U_2\over T_2} e^{2i\pi n({j+pU_1\over k})}
e^{2\pi kp T_2} \int_0^\infty
{d\tau_2\over \tau_2^{3/2+r}} e^{-{\pi T_2\over U_2}(k+{n U_2\over kT_2})^2
\tau_2}  e^{-\pi p^2 T_2 U_2/\tau_2}
\ee
The $\tau_2$ integration can now be done using the formula
\be
\int_0^\infty {dx\over x^{3/2+r}}  e^{-ax-b/x}
= \left(- {\partial\over \partial b}\right)^r \sqrt{\pi\over b}\;
 e^{-2\sqrt{ab}} 
\label{formula}
\ee
where  $a ={\pi T_2\over U_2}(k+{n U_2\over kT_2})^2$ and 
$b= \pi p^2 T_2 U_2$ are both proportional to the volume of the
compactification  torus.  
The leading term in the large-volume limit  is obtained when all
derivatives hit  the exponential in the above expression.
Using (\ref{formula}) we find
\be
I_{n,r} = {1\over k\vert p\vert T_2} \left( {k\over
 \vert p\vert U_2} \right)^r \; e^{2\pi k (p-\vert p\vert)T_2}
e^{2i\pi n [ {j+pU_1\over k} + i\vert p\vert {U_2\over k} ]}
\; \Bigl( 1 + {\rm o}({1\over T_2})\Bigr)
\label{formula1}
\ee
and plugging back into eq. ({\ref{3terms}) we obtain

\be
 {\cal I}^{het}_{inst} \simeq -{2 V^{(10)}\over 2^{10}\pi^6}
  \sum_{{0 \leq j<k} \atop { p > 0}}
      {1\over k p T_2}\;  e^{2\pi i Tpk}
        \; {\cal A}\left({j+p U\over k}\right) +\  {\rm c.c.} 
\label{inst}
\ee
This equality is exact for the holomorphic parts of the elliptic genus.
Correction terms have the form of an order-$r_{max}$ polynomial
in inverse powers of the volume, as we will discuss in a minute.

 Expression (\ref{inst}) has an elegant rewriting in terms of 
Hecke operators  $H_N$ \cite{Serre,DMVV}.
On  any modular form
 $\Phi_r(z)$ of weight $-2r$,
the action of a Hecke operator,  defined by  \cite{Serre}
\be
H_N[\Phi_r](z) = {1\over N^{2r+1}} 
\sum_{k,p>0\atop kp=N} \sum_{0\le j <k } k^{2r}\;
  \Phi_r\left(pz+j\over
k\right) \ ,
\label{hec}\ee
gives another modular form of the same  weight.
The Hecke operator is self-adjoint
with respect to the inner product defined by integration of modular
forms on a fundamental domain.
Using the above definition
 one finds
\be
{\cal I}^{het}_{inst} \simeq -{2 V^{(10)}\over 2^{10}\pi^6}  \sum_{N=1}^\infty
      {1\over  T_2}\;  e^{2\pi i NT}
        \; H_N[ {\cal A}](U) + {\rm c.c.} 
\label{inst1}
\ee 
In the above  form the result might be easier to compare with
a calculation based on the heterotic matrix string theory
 \cite{matrix2}.

\begin{figure}
\begin{center}
\leavevmode
\epsfbox{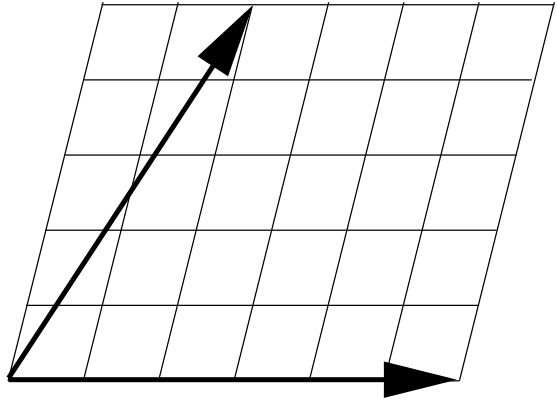}
\end{center}
\caption[]{Embedding of the lattice $\Gamma'$ (D1-brane) in the lattice $\Gamma$ (target space torus).}
\label{Lattic}\end{figure}

%%%%%%%%%%%%%%%%%%%%%%%%%%%%%%%%%%%%%%%%%%%%%%%%%%%%%%%%%%%%%%%%%%%%%

  Let us complete now the calculation, by taking into account the
sub-leading terms in the large-volume limit. 
Using eq. ({\ref{formula}) we can in fact evaluate explicitly
the integrals ({\ref{form2}). After some long but straightforward
algebra the correction terms can all be expressed in terms
of the induced moduli
\be
{\cal U}= {j+pU\over k}\ \ \ {\rm and } \ \ \ 
{\cal T}= kp T \ .
\ee     

\be
I_{n,1} \to I_{n,1}\times \left(\; 1 + {1\over {\cal T}_2}( n 
 {\cal U}_2  + {1\over 2\pi}) \right)\ ,
\ee
\be
I_{n,2} \to I_{n,2}\times \left(\;1 +  {1\over {\cal T}_2}( 2n
 {\cal U}_2 +  {3\over 2\pi})
+ {1\over {\cal T}_2^2}( n^2 {\cal U}_2^2 + {3n {\cal U}_2\over 2\pi}
+ {3\over 4\pi^2}) \right) \ .
\ee
These terms can be rewritten elegantly by using the 
 operator
\be
\square  \equiv {\cal U}_2^2 \partial_{\cal U} {\bar \partial}_{\cal U}
\ee
This is a modular invariant operator, which annihilates
holomorphic forms. 
The correction terms for all $r=0,1,2$ are summarized 
by the expression
\be
{\cal U}_2^{\ r} e^{- 2i\pi {\cal U}n}
\left( 1 + {1\over \pi {\cal T}_2}\square + {1\over 2}
{1\over \pi^2 {\cal T}_2^2 } (\square^2 - \square/2)\right)\
{\cal U}_2^{\; -r}  e^{ 2i\pi {\cal U}n} .
\ee
The instanton sum is modified accordingly to
\be
{\cal I}^{het}_{inst} = -{2 V^{(10)}\over 2^{10}\pi^6}  \sum_{instantons}
      {1\over  {\cal T}_2}\;  e^{2\pi i {\cal T}}
        \; \left( 1 + {1\over \pi {\cal T}_2}\square + {1\over 2}
{1\over \pi^2 {\cal T}_2^2 } (\square^2 - \square/2)\right)\
 {\cal A}({\cal U})\  + \  {\rm c.c.} \ .
\label{inst3}
\ee
One final rearrangement puts this to the form
\be
{\cal I}^{het}_{inst} = - {2 V^{(10)}\over 2^{10}\pi^6}  \sum_{instantons}
      {1\over  {\cal T}_2}\;  e^{2\pi i {\cal T}}
        \; \left( \sum_{s=0}^\infty {1\over s!}
 {1\over {\cal T}_2^s} (-iD)^s ({\cal U}_2^2 {\bar \partial}_{\cal U})^s
\right)\
 {\cal A}({\cal U})\  + \  {\rm c.c.} \ .
\label{inst4}
\ee 
where here $D$ is the covariant derivative, which acting on a modular
form $\Phi_r$ of weight $-2r$ gives a form of weight $-2r +2$,
\be
D \Phi_r = \left( {i\over\pi}\partial_{\cal U} 
 - {r\over \pi {\cal U}_2}\right)
\Phi_r \ .
\ee

The virtue of this last rewriting is that the $s$th operator in the
sum annihilates explicitly the first $s$ terms in the expansion
of the elliptic genus in powers of ${1\over {\cal U}_2}$.
From the general form of ${\cal A}$, eq. (\ref{exp}) we conclude
that 
 only the terms with $s\leq  2$ ($s\leq 1$) contribute in the
case of sixteen (eight) unbroken real supercharges.
 The modular-invariant
descendants of the genus, obtained  by applying the
 $s$th operator on ${\cal A}$,  
determine in fact  the corrections to  other dimension-eight
operators in the effective action. 
The full effective action can  be expressed in terms
of generalized holomorphic prepotentials, a result that we will
not develop further here.

\subsection{D-instanton Interpretation\label{dinst}}
\setcounter{equation}{0}

We would now like to understand the above result from the perspective
of type-I string theory. 
The world-sheet instantons on the heterotic side map  to
D-brane instantons, that is Euclidean trajectories of D-strings
wrapping non-trivially around the compactification torus. 
A  Euclidean  trajectory
described  by eq. (\ref{wrap}) defines a sublattice ($\Gamma^\prime$)
 of the compactification
lattice ($\Gamma$).
 If ${\bf e}_{i=1,2}$ are the two vectors spanning $\Gamma$, then
$\Gamma^\prime$ is spanned by the vectors
${\bf e^\prime}_{i} = M_{ji} {\bf e_j}$ (figure \ref{Lattic}).
Under a change of basis for $\Gamma$ ($\Gamma^\prime$) the matrix
$M$ transforms by  left  (right) multiplication with
the appropriate elements  of SL(2,{\bf Z}). Using reparametrizations
of the  world-sheet we can thus bring the basis
 ${\bf e^\prime_i}$ into the canonical form, eq. (\ref{nondeg}),
 as described in
the previous section 
 (see also  figure \ref{Lattic}).

 Now the key remark is that on  the heterotic  world-sheet 
we have an  induced  complex structure and K{\"a}hler modulus,  
which for positive $p$ are given by
\be
{\cal U}= {j+pU\over k}\ \ \ {\rm and } \ \ \ 
{\cal T}= kp T \ .
\ee
For negative $p$'s,  describing  anti-instantons, we must take the
absolute value of $p$ and complex conjugate these  expressions.
One can check these facts by inspection of figure \ref{Lattic},
 or by computing
explicitly the
 pull-backs of the metric and
antisymmetric tensor field,
\be
\hat G_{\alpha\beta}=G_{\m\n}\partial_\alpha 
X^{\mu}\partial_\beta  X^{\nu}\,\,\,,\,\,\,
\hat B_{\alpha\beta}=B_{\m\n}\partial_\alpha
X^{\mu}\partial_\beta  X^{\nu} \ .
\label{31}
\ee
Notice that $N=kp$ is the total number of times the world-sheet
wraps around the compactification torus. 
In terms of induced moduli the instanton sum (\ref{inst})
 takes the form
\be
{\cal I}_{inst} \simeq -{2 V^{(10)}\over 2^{10}\pi^6}  \sum_{instantons}
      {1\over  {\cal T}_2}\;  e^{2\pi i {\cal T}}
        \;  {\cal A}({\cal U}) +  {\rm c.c.} \ .
\label{inst2}
\ee

\begin{figure}
\begin{center}
\leavevmode
\epsfbox{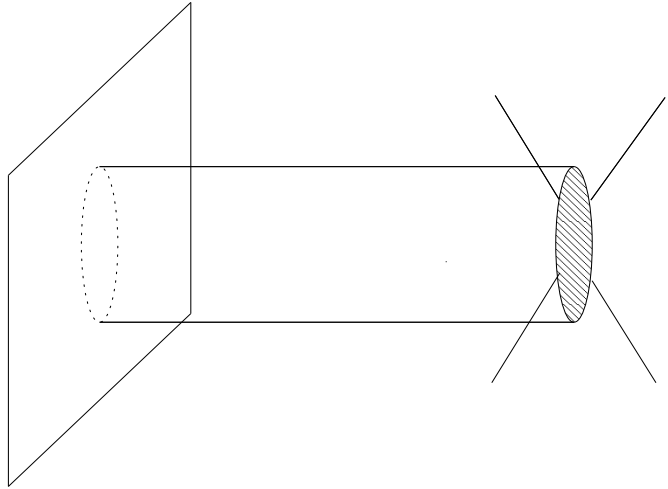}
\end{center}
\caption[]{A D1-brane instanton correction to $tr F^4$.}
\label{f5}\end{figure}

The various terms of this expression have a simple interpretation
on  the type-I side. The action of a wrapped D-string is
 \cite{tasi}
\be
S_{\rm D-string}=
{1 \over 2\pi  \alpha'\lambda_I}\int d^2\sigma \sqrt{|{\rm det}
\hat G_I |}-{ i\over 2\pi \alpha'}\int \hat B_I
\label{30}
\ee
where $B_I$ is the type-I 2-form coming from the RR sector.
Using the heterotic/type-I map
\be
{\cal T}_2^{het} = {\cal T}_2^I/\lambda_I \ , \ \  B^{het} = B^I
\ee
and the fact that the world-sheet area of the D-string 
is $4\pi^2 {\cal T}_2^I$, we see that the exponential of this
Nambu-Goto action reproduces exactly the  exponential in the instanton
sum, eq. (\ref{inst2}). The inverse factor of the volume
comes from the integration of the longitudinal translation zero modes.
Finally the elliptic genus of the D-brane complex structure, 
should come from the  functional integration  over the
(second quantized) string fields in the instanton 
background. A typical  diagram contributing to the ${\cal F}^4$
coupling is shown in figure 3.
 For the purely holomorphic pieces of the elliptic
genus the result is topological, so it should be  expected to coincide
with the heterotic
 $\sigma$-model calculation of refs. \cite{Schellekens,Lerche,Windey,ellwit}.
Put differently, massive string modes and higher-order terms in
the effective D-string action are expected  to play no role in the 
calculation.

 From the type-I point of view expression (\ref{inst}) is,
however, still somewhat unnatural.
 The three configurations of figure \ref{f6} correspond
to the same (singular) effective-field-theory solution, characterized
by two units of the appropriate Ramond-Ramond charge. Why then
should we count them as distinct saddle points?
 Furthermore, the fluctuations
of the double D-brane  are not described by the usual heterotic
$\sigma$-model, but by its (non-abelian)  $2\times 2$-matrix
generalization \cite{matrix,matrix2},
which is the low-energy limit of an open string theory.
Why should then the result be  proportional to the conventional
elliptic genus?

In order to answer these questions
 it is convenient to put the
effective action (\ref{inst}) in the more elegant form
\begin{equation}
 {\cal I}^{\rm inst} =
- { V^{(8)}\over 2^{8}\pi^4}
\sum_{N=1}^\infty   \;  e^{2\pi i N T}
 {\cal H}_N  {\hat {\cal A}}(U)\ + {\rm c.c.},
\label{Hecke}
\end{equation}
with
\begin{equation}
{\cal H}_N {\hat {\cal A}}(U) \equiv
{1\over N} \sum_{kp = N \atop 0\le j <k}
 {\hat {\cal A}}\left({j+p U\over k}\right) \ .
\end{equation}
We have just seen the geometric interpretation of the Hecke operators
in terms of inequivalent N-fold wrappings of the torus by the
(heterotic) world-sheet. We will now describe an
 alternative interpretation,
more appropriate on the type-I side, in terms of the moduli space of
instantons \cite{bv,bk2}. The key 
 will be  to treat this moduli space as a
symmetric orbifold \cite{c,Cum,DMVV}, an idea that is more familiar
in the context  of black-hole state counting \cite{BH}.

\begin{figure}
\begin{center}
\leavevmode
\epsfbox{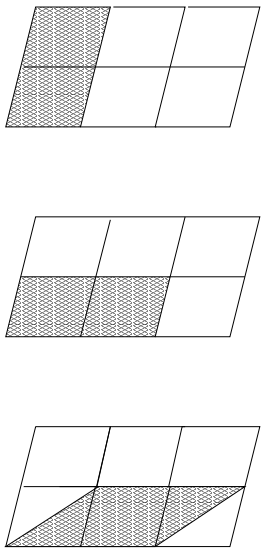}
\end{center}
\caption[]{The three distinct ways in which the (shaded) 
heterotic world-sheet can wrap
twice around the target-space torus.}
\label{f6}\end{figure}

The low-energy
 fluctuations around  a configuration of
$N$ instantonic D-branes  are  described by a 
heterotic matrix $\sigma$-model, with local
SO(N) symmetry on the world-sheet \cite{matrix,matrix2}.
The
 coupling of  a
constant target-space  background field reads
$$
\delta I_{\sigma}  \  \propto\ 
 \int \;   F_{ij}^\alpha T_\alpha^{rs}\ 
\lambda_r^{\; T} \ 
\Bigl[  X^i {\bar D}  X^j - {1\over 8}  S^\dagger 
 \gamma^{ij} S \Bigr] \lambda_s \ .
$$
Under the SO(N) gauge symmetry the
supercoordinates $X^i$ and $S^{\dot a}$ are symmetric
matrices,  the  current-algebra fermions
$\lambda_r$  are vectors, and ${\bar D}$ is the antiholomorphic
covariant derivative.
We are interested in the functional integral of
 this $\sigma$-model,
with four insertions of $\delta I_{\sigma}$. Notice that 
 contributions of massive string modes are expected to cancel out
for this special amplitude, justifying  the reduction of the
calculation to the matrix model.

The moduli space of this multi-instanton has  a Coulomb branch 
along which the $X^i$ have diagonal
expectation values. In the type $I^\prime$ language these label
the  positions of  N D0-particles on  the orientifold plane
\footnote{There is also a Higgs branch, corresponding to the motion
of mirror pairs of D0-particles off the orientifold plane.
Because of the SO(N)-gaugino zero modes, this part of the moduli
space does  not contribute.}.
 At a generic point in this moduli space there are 8N fermionic
zero modes, corresponding to the diagonal components of the
matrices $S^{\dot a}$. Since only eight of them
can  be absorbed by the  insertions of the vertex $\delta I_{\sigma}$,
 one would naively conclude that the 
sectors $N>1$ do not contribute.  This is wrong because 
of the residual gauge symmetry that permutes the positions of the
D-branes. {\it The 
moduli space is thus a symmetric  orbifold  and there
are potential  contributions from its fixed points}.

  Let us illustrate how this works in  the case
 of two  instantons. The massless
fluctuations of the double D-brane are described by a conformal
field theory with target space ${\cal M}\times {\cal M}/Z_2$,
where ${\cal M}$ is the (transverse)
 target space of the heterotic string,
and $Z_2$ is the exchange symmetry. There are four contributions to the
amplitude, corresponding to the four 
boundary conditions on the torus. The untwisted sector has $2\times 8$
fermionic zero modes and does not contribute.
The contribution of
the remaining three sectors is proportional to
$$
{\hat {\cal A}}(2U) + {\hat {\cal A}}\Bigl({U\over 2}\Bigr)+
{\hat {\cal A}}\Bigl({U+1\over 2}\Bigr)\ ,
$$
as can be shown using  standard $Z_2$-orbifold techniques.
This is precisely the action of the Hecke operator ${\cal H}_2$,
corresponding to the sum over the three surfaces of figure 1. 
  The overall coefficient also
checks, including the orbifold normalization of ${1\over 2}$,  and
 the simple factor of the transverse volume
 characteristic of
the twisted-sector contributions.

  The generalization to any $N$ is straightforward. The target
space is now the symmetric orbifold 
$$
\underbrace{ {\cal M} \times ... \times {\cal M}}_{\rm N\  times}
\bigg/S_N 
$$
The non-vanishing contributions
to the amplitude come from those boundary conditions for which 
only the trace part of $S^{\dot a}$ is (doubly) periodic on the torus.
Up to a common overall normalization, the
 result is given by  ${\cal H}_N {\hat{\cal A}}(U)$,
which is
 the matrix-model generalization of  ${\hat{\cal A}}$.
The non-perturbative type-I effective action is obtained by 
 summing  over all N,  as in  expression 
(\ref{Hecke}).

\boldmath
\subsection{Heterotic/Type-I duality  in $D<8$\label{d8}}
\setcounter{equation}{0}
\unboldmath

We will now discuss heterotic thresholds in toroidal compactifications
to $D<8$. As we argued earlier, if $D>4$ then the heterotic result is
still one-loop only and can be evaluated.
Using heterotic/type-I duality we find again the non-perturbative
type-I corrections and we show that their corresponding D1-brane
interpretation
is in agreement with the D1-brane rules given in Section \ref{dinst}.

Our starting point is the general form of the one-loop thresholds
\be
{\cal I}^{\rm het}_{D}  =
-{\cal N}_D
  \int_{ {\cal  F}}{d^2\tau \over
\tau_2^2}\; (\tau_2^{d/2}   \Gamma_{d,d}(G,B ))\
{\cal A}(\tau)
\co \label{ddh}
\ee
where the $D + d =10$ and the $d$-dimensional lattice sum $\Ga_{d,d}$
is given by
\be
\Ga_{d,d}(G,B)  = {\sqrt{G} \over \t_2^{d/2} }
\sum_{m^i,n^i \in \Bbb Z }
\exp\left[-{\pi\over \tau_2}(G+B)_{ij }
(m^{i}+ n^{i} \t )(m^{j}+ n^{j} \bar\t )\right]
\co \label{ddt} \ee
where $G$ and $B$ are the $d$-dimensional metric and antisymmetric
tensor
respectively. 

The corresponding integral (\ref{ddh}) can be evaluated again, using the
method
of orbits. We refer to \cite{ko} for the main steps, and quote
here only the result of the non-degenerate orbit, which comprises the
type-I instantonic contributions:
\be
 {\cal I}_{\rm inst}=-2 {\cal N}_D
\sum_{s=0}^{\n_{\rm max}}
 \left(  { 3 \over 2 \p} \right)^s
 \sum_{m,n}' { \sqrt{G} \over (T^{m,n}_2)^{s+1} }
 e^{2 \pi i T^{m,n} }
\cA_s( U^{m,n})
\label{gic} \ee
Here, the induced K\"ahler and complex structure moduli are given by
\bs
\be
T^{m,n} =  m B n + i \sqrt{ (m G m) (n G n  ) - (m Gn)^2 }
\;\;\;\;\;\;\;\;\;\;\;\;\;\;\;\;\; \ee
\be
U^{m,n} = \left( - m G n + i \sqrt{ (m G m) (n G n  ) - (m Gn)^2 } \;
\right)
/ n G n
\ee
\label{utm} \es
and the $\sum_{m,n}'$ is over the non-degenerate orbits, which are
parametrized
by the following integer-valued $2\ti d$ matrices
\bs
\be
\mbox{non-degenerate orbit} : \;\;\;\; A^T=
 \left( \matrix{n_1 &  \ldots & n_k & 0 &  \ldots &   0  \cr
 m_1 &  \ldots & m_k & m_{k+1} &  \ldots  & m_d   \cr } \right)
\;\;\;\;\;\;\;\;\;\;\;\;\;\;\;\;\;\;
\;\;\;\;\;\;\;\;\;\;\;\;\;\;
\ee
\be
\;\;\;\;\;\;\;\;\;\;\;\;\;\;\;\;\;
\;\;\;\;\;\;\;\;\;\;\;\;\;\;\;\;\;
 1 \leq k < d   \sp n_k > m_k \geq 0 \sp (m_{k+1},\ldots,m_d)  \neq
(0,\ldots,0)
\pe  \ee
\label{ndm}
\es
Note that for $d=2$ the general result (\ref{gic}) reduces to the one
given in
(\ref{inst}).

Turning to the D1-brane interpretation of the result, we first wish
to establish that the exponential factor $ e^{2 \pi i T^{m,n} } $
agrees
with the classical action of a D1-brane. The map that describes the
wrapping of the D1-brane world-sheet around a 2-cycle in the $d$-torus
is
\be
X^i = n_i \s_1 + m_i \s_2  \sp i = 1 \ldots d
\co \label{dtm}
\ee
where $X^i$ are the coordinates on $T^d$ and $\s_{1,2}$ the D1-brane
coordinates. We observe that modular transformations on the D1-brane
coordinates act on the matrix $A$ that enters (\ref{dtm})
\be
 A =
 \left( \matrix{n_1 &  m_1  \cr
\vdots & \vdots \cr
 n_d  &  m_d \cr} \right)
\ee
by right SL(2,\Z) transformations, which forces us to pick the
representative
configurations described by the matrices in (\ref{ndm}).

In terms of the matrix $M_I^i =(A_i^I)^T =  (n^i,m^i)$, $I=1,2$, we
see that
the induced metric and antisymmetric tensor fields are
\be
\hat{G}_{IJ} = M_I^i G_{ij} M_J^j
\sp \hat{B}_{IJ} = M_I^i B_{ij} M_J^j
\pe \label{inf}
\ee
In particular, going through the same steps as in Section \ref{dinst}, we find
from the D1-brane classical action (\ref{30}) and (\ref{utm}),
(\ref{inf})
that $e^{-S_{\rm class}}$ precisely reduces to the
exponential factor $ e^{2 \pi i T^{m,n} }   $,
which is to be summed over the ranges indicated in (\ref{ndm}).
We also note that we correctly observe the overall factor
$\sqrt{G}/\sqrt{\hat{G}} = \sqrt{G}/T^{m,n}_2 $. Moreover, the
fluctuation
determinant is evaluated at the induced modulus $U^{m,n}$ of the
wrapped D1-brane.

This establishes the claim that the D1-brane rules in $D<8$ are
consistent
with those obtained for $D=8$. In summary, we have found the
intuitively
expected result:
The situation is as in eight dimensions with the difference that now the D1 
world-volume can wrap in many more ways on submanifolds of $T^{10-D}$.

In the eight-dimensional case it was shown \cite{bk2,ko} that differential
equations satisfied by the (2,2) toroidal lattice sum translate into
recursion relations for the thresholds, which can be solved in terms of
holomorphic prepotentials.
There is a generalization of such equations for the $(d,d)$ toroidal lattice
sum.

It was noted in Refs. \cite{KK,KK5} that the toroidal partition function
$\Ga_{d,d}(G,B;\t) $ satisfies the following differential
equation:
\be
\left[
\left( \sum_{i \leq j} G_{ij} { \pa \over \pa G_{ij} } +
{ 1- d \over 2} \right)^2
+ \frac{1}{2} \sum_{ijkl} G_{ik} G_{jl} {\pa^2 \over \pa B_{ij} \pa B_{kl} }
 -\frac{1}{4}
 - 4 \t_2^2 {\pa^2 \over \pa \t \pa \bar{\t} }
 \right] \Ga_{d,d}(G,B;\t)  = 0
\label{de1} \ee
which in the case $d=2$ reduces to
\be
\left[  T_2^2 \pa_T  \pa_{\bar{T}}
 -  \t_2^2  \pa_\t \pa_{\bar{\t}}
 \right] \Ga_{2,2}(T,U;\t)  = 0 \pe
\label{2de} \ee
However, the general differential equation in (\ref{de1}) is not
invariant under
the full $O(d,d,Z)$ duality group. It  may be verified that
it is invariant under integer $B$ shifts and $SL(d,Z)$ basis changes,
but there is no invariance under the
 remaining generators of the duality group, which
are the inversion and factorized duality. The latter
two transformations act on the matrix $E \equiv G +B$ as follows:
\be
E \ra E^{-1} \sp E \ra [(1-e_i)E + e_i][ e_i E + (1-e_i)] ^{-1}
\sp (e_i)_{k,l} = \d_{ik} \d_{il}
\pe \ee
For example, in the  $d=2$ case
the factorized dualities correspond to $ T \ra U$ and $T \ra 1/U$ for $i=1$
and 2 respectively, which do not leave
the differential equation in (\ref{2de}) invariant.

This implies that there must be further constraints on $\Ga_{d,d}$
generalizing the $d=2$ relation
\be
\left[  T_2^2 \pa_T  \pa_{\bar{T}}
-  U_2^2 \pa_U  \pa_{\bar{U}} \right]  \Ga_{2,2}(T,U;\t)  = 0
\pe \label{utd} \ee
To find the generalization of this relation we note that there
is another O(d,d,\Z)-invariant differential equation
on the lattice sum, which
reads
\be
\left[  \sum_{ijkl}
 G_{ik} G_{jl} {\pa^2 \over \pa E_{ij} \pa E_{kl} }
+   \sum_{ij} G_{ij} { \pa \over \pa E_{ij} }
+\frac{1}{4} d (d-2) -4 \t_2^2 \pa_{\t} \pa_{\bar{\t}} \right]
\Ga_{d,d}(E;\t)  = 0
\pe \label{de2} \ee
This is in fact the O(d,d) Laplacian \cite{eisen}.
As a consequence we find that the difference between (\ref{de1}) and
(\ref{de2}) is the differential equation,
\be
\left[ \sum_{ijkl} (G_{ij} G_{kl} - G_{jk} G_{il})
{\pa^2 \over \pa E_{ij}  \pa E_{kl} }
+ (1-d)\sum_{ij} G_{ij}  {\pa \over \pa E_{ij}  }\right]
\Ga_{d,d}(E;\t)  = 0
\co \label{de3} \ee
which, for $d=2$, turns out to precisely reduce to (\ref{utd}).

Clearly (\ref{de3})
is not invariant under the duality group, since (as (\ref{de1}))
the inversion and factorized duality are broken, but  these transformations
should be used to form a complete irreducible set of differential equations.

The threshold as well as the associated differential equations were analysed 
further in 
\cite{eisen}. There, the perturbative threshold was conjectured to be equal 
to the s=1
Eisenstein series of the spinor of O(d,d).

It is an open problem to 
define the analog of prepotentials in the lower-dimensional case.

\section{N=4 $\Rr^2$ couplings and five-brane instantons\label{r2}}

In this section we will discuss $\Rr^2$ effective couplings in theories with N=4 supersymmetry. In particular we will analyse type II (2,2) vacua as well
as the dual heterotic vacua.
The prototype (2,2) vacuum is type II theory compactified on K3.
The dual vacuum is heterotic theory compactified on $T^4$. We will analyse this dual
pair and follow it also in four dimensions.

\subsection{General remarks}

As mentioned in a earlier section, contributions to $\Rr^2$ couplings 
depend on the type of $N=4$ vacua
we are considering: ($2,2$) vacua, where two supersymmetries come
from the left-movers and two from the right-movers, or ($4,0$) vacua,
where all four supersymmetries come from the left-movers only.
All heterotic ground-states with $N=4$ supersymmetry are of the ($4,0$)
type, but ($4,0$) type~II vacua  can also be constructed
\cite{fk}. In that case, the axion--dilaton corresponds to the
complex scalar in the gravitational multiplet in four dimensions
and, as such,
takes values in an SU(1,1)/U(1) coset space,
while the other scalars
form an SO(6,$N_V)\Big/\Big($SO(6)$\times $SO($N_V)\Big)$ manifold, where
$N_V$ is the
number of vector multiplets in four dimensions.
On the other hand
($2,2$) vacua only exist in the type II theory and have a different
structure: the dilaton is now part of the $SO(6,N_V)\Big/\Big(SO(6)\ti
SO(N_V)\Big)$
manifold, while the SU(1,1)/U(1) coset is spanned by a perturbative
modulus. Duality always maps a ($2,2$) ground-state to a ($4,0$) ground
state \cite{ht}.
We shall argue that $\Rr^2$ couplings are exactly given
by their one-loop result in all ($2,2$) vacua. Translated
into the dual ($4,0$) theory, the exact $\Rr^2$ coupling
now appears to arise from non-perturbative effects,
which can be identified with NS5-brane instantons \cite{HM,6}.

At tree level, $\Rr^2$ terms can be obtained directly from the relevant
ten-dimensional calculations (see \cite{slo}) upon compactification
on the appropriate manifold, K3, K3$\times T^2$ or $T^6$.
They turn out to be non-zero in ($4,0$) ground-states  (heterotic or
type
II)
and zero for ($2,2$) ground-states. They may a priori also receive
higher-loop perturbative corrections, but
($4,0$) ground-states appear to have no perturbative corrections at
all,
while the perturbative corrections in ($2,2$) vacua are expected
to come only
from one loop owing to the presence of extended
supersymmetry.

These terms are related by
supersymmetry to eight-fermion couplings.
As such they may receive non-perturbative corrections
from instantons having not more than 8 fermionic zero-modes.
This rules out generic instanton configurations, which break
all of the 16 supersymmetric charges and therefore
possess at least 16 zero-modes. However, there exist
particular configurations that preserve one half of the
supersymmetries (this is the only possibility in six dimensions
where there only two supercharges), thereby possessing 8
fermionic zero-modes\footnote{Instantons with less than 8
zero-modes do not exist, in agreement with the absence of
corrections to the two-derivative or four-fermion action.}.
These configurations correspond to
the various $p$-brane configurations of the original ten-dimensional
theory.

All superstrings in ten dimensions have in common the NS
5-brane that couples to the dual of the NS--NS antisymmetric tensor and
breaks
half of the ten-dimensional supersymmetry.
Type II superstrings also have D $p$-branes that are charged under the
various R--R forms
and their duals:
$p=0,2,4,6,8$ for type IIA theory,
$p=-1,1,3,5,7$ for type IIB. Obviously,
D-branes are absent from heterotic ground-states. 
The only instanton configuration for such vacua
is therefore the NS5-brane, which only starts to contribute for
dimensions
less than or equal to four.

In ($2,2$) models the situation is a bit more involved.
Let us consider first the type IIA or IIB string compactified on K3 to
six dimensions.
Since K3 is four-dimensional, only branes with $p+1\leq 4$ need be
considered as instantons.
Wrapped in a generic fashion around  submanifolds of K3
they
break all
supersymmetries and thus do not contribute, in our calculation.
There are, however, supersymmetric $0, 2$ and 4 cycles in K3.
The relevant instantons will then have $p+1=0,2,4$, found only in type
IIB.
Thus in type IIA theory we do not expect any instanton corrections.
In type IIB theory, all scalar fields span an
SO(5,21)$\Big/\Big($SO(5)$\times $SO(21)$\Big)$ coset
space.
The perturbative $T$-duality symmetry O(4,20,\Z) combines with the
SL(2,\Z) symmetry in ten dimensions into an O(5,21,\Z)
U-duality symmetry group.
The exact non-perturbative threshold should therefore be an
O(5,21,\Z)-invariant function of the moduli and, as argued in
\cite{kp1,kp2,neuc,eisen}, it
can be written as linear
combinations
of the Eisenstein--Poincar\'e series. However, all such series
have distinct and non-zero perturbative terms when expanded in terms of
any
modulus, in disagreement with the fact that all perturbative
corrections
should vanish.
We thus conclude that the $\Rr^2$ threshold is  non-perturbatively zero
also in type IIB on K3.

There is an independent argument pointing to the same result.
Consider compactifying type IIA, B on K3$ \times S^1$.
Then IIA and IIB are related by
inverting the circle radius.
{}From the type IIA point of view there are now potential instanton
corrections
from the $p=0,2,4$-branes wrapping around a $0,2,4$ K3 cycle
times $S^1$.
However, on the heterotic side we are still in a dimension larger than
four
so we still have no perturbative or non-perturbative corrections.
This implies that the contribution of the IIA instantons still
vanishes,
as it does for the IIB instantons, which are just the same as the
six-dimensional ones. The instanton contributions in six dimensions
thus also have to vanish.

Compactifying further to four dimensions on an extra circle,
the scalar manifold becomes
SU(1,1)/U(1)$ \times  $SO(6,22)$\Big/\Big($SO(6)$\times $SO(22)$\Big)$
and the duality group
SL(2,\Z)$\times$O(6,22,\Z).
The instanton contributions
can come from 5-branes wrapped around K3$\times T^2$ as well as,
in type IIB, from D3-branes wrapped around $T^2$ times a
K3 2-cycle, and $(p,q)$
D1-branes wrapped around $T^2$.
The D1-brane contribution is zero since it is related via SL(2,\Z)
duality to that of the fundamental string world-sheet
instantons, which vanish
from the one-loop
result\footnote{This is equivalent to the statement that in IIB
the one-loop threshold only depends on the complex structure $U$
of the torus. This no longer holds for other
thresholds such as $\nabla H \nabla H$ and it is found \cite{6}
that those are non-perturbatively corrected even in type II.}.
All other instanton corrections depend non-trivially on the $O(6,22)$
moduli.
Again, it is  expected that an $O(6,22)$-invariant result would imply
perturbative corrections depending on the $O(6,22) $ moduli, which are
absent, as we
will show. Therefore,  we again obtain that the
non-perturbative corrections vanish in IIB, and also in IIA.
This can also be argued via type-II/heterotic/type-I triality.
On the heterotic and type-I side these corrections come from the
5-brane
wrapped on $T^6$. The world-volume action of the D5-brane in type II
theory
is known (and will be calculated further on).
Wrapping it around $T^6$ and translating to heterotic
variables produces a result depending only on the $S$ field.
Thus on the heterotic side we do not expect $O(6,22)$-dependent
corrections, and therefore no instanton contributions in type II.

The upshot of the above discussion is that, in $(2,2)$ models, various
dualities imply that
on the type \-II side instanton corrections to $\Rr^2$ terms are absent in
six, five and four dimensions.

\setcounter{equation}{0}
\subsection{One-loop corrections in six-dimensional
type IIA and IIB theories\label{6d}}

In this section, we compute the one-loop four-derivative terms in the
effective action for type IIA and IIB theory compactified to six
dimensions on
the K3 manifold. We will work in the $Z_2$ orbifold limit of K3 in
order to be explicit but,  as we will show, the result will be valid
for
all values of the K3 moduli.
To compute the massless spectrum we need the following geometric data
of K3: the Einstein metric on K3 is parametrized by 58 scalars, and
the non-zero Betti numbers are $b_0=b_4=1$ and $b_2=22$.
 Out of the 22 two-forms, 3 are
self-dual, while the remaining 19 are anti-self-dual. At the $T^4/Z_2$
orbifold point of K3, those correspond to the $3+3$ $Z_2$-even
two-forms $\rd x^i \w \rd x^j$ and to 16 anti-self-dual two-forms
supported by
the two-sphere that blows up each of 16 fixed points.
With this in mind, it is easy to derive the massless spectrum:

{\sl Type IIA.} The ten-dimensional bosonic massless spectrum
consists of the  NS--NS fields
$G_{MN}$, $B_{MN}$, $\Phi$ and of the R--R three-form and one-form
potentials
$A_{MNR}$ and $A_{M}$. Compactification on K3 then gives in the
NS--NS
sector
$G_{\m \n}$ and 58 scalars,  $B_{\m \n}$ and 22 scalars, and the
dilaton
$\Phi$; in the R--R sector we have $A_{\m \n \r}$ and 22 vectors in
addition
to $A_{\m}$.
In six dimensions, $A_{\m \n \r}$ can be dualized into a vector, so
all in all the bosonic fields comprise a graviton, 1 antisymmetric
two-form tensor, 24 $U(1)$ vectors
and 81 scalars.  Hence, we end up with the following supermultiplets of
six-dimensional ($1,1$) (non-chiral) supersymmetry:
\be
\mbox{1 supergravity multiplet} \sp
\mbox{20 vector multiplets}\, ,
\ee
where we recall that:
\nl
-- the ($1,1$) supergravity multiplet comprises a  graviton, 2 Weyl
gravitinos of
opposite chirality, 4 vectors, 4 Weyl spinors of opposite chirality,
1 antisymmetric tensor, 1 real scalar;
\nl
-- a vector multiplet comprises 1 vector,  2 Weyl spinors of opposite
chirality,
4 scalars. \nl
The scalars parametrize R$^{+}\ti$SO(4,20)$\Big/\Big($SO(4)$ \ti$
SO(20)$\Big)$, where the first
factor corresponds to the dilaton up to a global  O(4,20,\Z)
$T$-duality identification.

{\sl Type IIB.} The ten-dimensional massless bosonic spectrum
consists of the NS--NS fields $G_{MN}$, $B_{MN}$, $\Phi$, and the
self-dual
four-form $A^+_{MNRS}$, the two-form $A_{MN}$ and the zero-form  $A$
from the
R--R sector. Compactification on K3
then gives in the NS--NS sector the same as for type IIA. In the R--R
sector, we obtain respectively $A^+_{\m \n \r\s}$ (which is not
physical),
22 $B^{\rm R-R}_{\m \n}$
(of which 19 anti-self-dual and 3 self-dual) and 1 scalar,
$A_{\m \n}$ and 22 scalars, and the scalar $A$ itself.
If we decompose both $B_{\m \n}$ and $A_{\m \n}$ into a self-dual and
an
anti-self-dual part, the bosonic content comprises a graviton,
5 self-dual and 21 anti-self-dual antisymmetric tensors and 105
scalars.
Hence, we end up with the following six-dimensional ($2,0$) (chiral)
supermultiplets:
\be
\mbox{1 supergravity multiplet} \sp
\mbox{21  tensor multiplets}\, ,
\ee
where we recall that:
\nl
-- the ($2,0$) supergravity mulitplet comprises a graviton, 5 self-dual
antisymmetric
tensors, 2 left Weyl gravitinos, 2 Weyl fermions;
\nl
-- a ($2,0$) tensor multiplet comprises  1 anti-self-dual antisymmetric
tensor, 5
scalars, 2 Weyl fermions of chirality opposite to that of the
gravitinos.
\nl
The scalars including the dilaton parametrize the coset space
SO(5,21)$\Big/\Big($SO(5)$ \ti $SO(21)$\Big)$, and the
low-energy supergravity has  a global O(5,21,\R) symmetry
\cite{romans}.
The O(5,21,\Z) subgroup is the non-perturbative U-duality symmetry \cite{ht}.

We will consider the three graviton or antisymmetric tensor 
 scattering amplitude at one loop.
We are interested in the piece quartic in momenta of the
 three-point function (since the terms we are after are four-derivative terms
 ):
\be
{\cal I} = \e_{1\m\n} \e_{2\k \l}
\e_{3\r \s}
\ifdd \int
\prod_{i=1}^3  {\rd^2 z_i  \over \p}
\left\langle V^{\m \n}(p_1,\bz_1,z_1)
 V^{\k \l}(p_2,\bz_2,z_2)
V^{\r \s} (p_3,\bz_3,z_3)
\right\rangle\, .
\label{64}
\ee
Here the space-time indices run over $\m = 0, \ldots ,5$
(see \cite{6} for conventions), and the
vertex operators in the 0-picture are
\be
V^{\m \n} (p,\bz,z) =
\left(\bpa X^\m (\bz,z) + i p \cdot \bps (\bz) \bps^\m (\bz) \right)
\Big(\pa X^\n (\bz,z) + i p \cdot \ps (z) \ps^\n (z)\Big)
{\rm e}^{i p \cdot X (\bz,z) }\, ,
\label{vop}
\ee
where the polarization tensor $\e_{\m \n}$ is symmetric
traceless for
a graviton ($\r\equiv 1$) and antisymmetric for an antisymmetric
two-form
gauge field
($\r\equiv -1$).

Altogether the physical conditions are
\be
\e_{\m \n} = \r \e_{\n \m} \sp
 p^{\m} \e_{\m \n} = 0 \sp
 p^{\m} p_{\m} =0 \sp
 p_1+p_2+p_3 =0\, .
\label{66}
\ee
Note that they imply $p_i \cdot p_j=0$ for all $i,j$. Were the $p_i$'s
real
and
the metric Minkowskian, this would indicate that the momenta are in
fact
collinear, and all three-point amplitudes would vanish due to
kinematics.  This can be
evaded by going to complex momenta in Euclidean space.

The expression (\ref{vop}) gives the form for all the vertex operators
when we take the even spin structure both on the left and the right.
When one spin structure (say left) is odd, though,
the presence of a conformal Killing
spinor together with a world-sheet gravitino zero-mode requires
one of the vertex operators (say the last one) be converted to the
$-1$-picture on the left
\be
V^{\m \n} (p,\bz,z) =
\left(\bpa X^\m (\bz,z) + i p \cdot \bps (\bz) \bps^\m (\bz) \right)
  \ps^\n (z) {\rm e}^{i p \cdot X (\bz,z) }\, ,
\label{mop} \ee
and a left-moving supercurrent
\be
G_F^{\vphantom{int}} =  \pa X^\g \ps_\g + G_F^{\rm int}
\label{68}
\ee
be  inserted at an arbitrary point on the world-sheet \cite{gm}.

There are four possible spin-structure combinations to consider,
which can be grouped in two pairs according to whether they describe
CP-even or CP-odd couplings,
\be
\mbox{CP-even:}\, \cases{\bar{e}{\rm-}e \cr \bar{o}{\rm-}o \cr }
\quad
\mbox{CP-odd:}\,  \cases{\bar{e}{\rm-}o \cr \bar{o}{\rm-}e \, ,\cr }
\label{cpc}
\ee
where we denote $e$ ($o$) the even (odd) spin structure on the left
and the barred analogues for those on the right.

The low-energy action can then be determined by finding
Lorentz-invariant terms that yield the same vertices on-shell.
Depending on the polarization of the incoming particles, the
string amplitude can be reproduced by the following terms in
the effective action (see \cite{6} for more details):
\be
\label{R2vert}
\Rr^2
\equiv
R_{\m\n\r\s} R^{\m\n\r\s}
\ee
\be
\na H \na H
\equiv
\na_\m H_{\n\r\s} \na^\m H^{\n\r\s}
\ee\be
B \w R \w R
\equiv
\e^{\m\n\k\l\r\s}_{\phl} B_{\m \n}^{\phl}
R_{\k\l}^{\hphantom{\k\l}\a\b}
R_{\r\s\a\b}^{\phl}
\ee\be
B \w \na H \w \na H
\equiv
\e^{\m \n \k \l \r \s}_{\phl} B_{\m \n}^{\phl} \na_{\k}^{\phl}
H_{\l}^{\hphantom{\l}\a \b}
 \na_{\r}^{\phl} H_{ \s \a \b}^{\phl}
\ee
\be
H \w H \w R
\equiv
\e^{\m \n \k \l \r \s}_{\phl} H_{\m \n \k}^{\phl}
H_{\l}^{\hphantom{\l}\a \b}
 R_{ \r \s \a \b}^{\phl}
\ee
$H_{\m\n\r}=\pa_{\m} B_{\n\r} + \pa_{\r} B_{\m\n} + \pa_{\n} B_{\r\m}$
is the field strength of the two-form potential, and the left-hand side
defines a short-hand notation for the corresponding term (in agreement
with standard notation up to factors of $\sqrt{-g}$).

Note that other four-derivative terms such as squared Ricci tensor
or squared scalar curvature do not contribute at three-graviton
scattering in traceless gauge, so that their coefficient cannot
be fixed at this order. That this remains true at four-graviton
scattering was proved in~\cite{fotw}; it can be seen as a consequence
of the field redefinition freedom
$g_{\m\n} \rightarrow g_{\m\n} + a R_{\m\n} + b R g_{\m\n}$,
which generates $R^2$ and $R_{\m\n}R^{\m\n}$ couplings from
the variation of the Einstein term.
Similarly, the coupling of two antisymmetric tensors and one graviton
could as well be reproduced by a variety of $RHH$ terms, equivalent
under field redefinitions.

We can do the calculation at the $Z_2$ orbifold limit of K3 \cite{6}.
We will argue below that the result is valid for arbitrary K3 moduli.

\begin{itemize}
\item{}
The ${\bar{e}{\rm-}e}$ sector manifestly receives $O(p^4)$
contributions from contractions of four fermi- ons on both sides,
and the resulting terms in the effective action are
\be
{\cal I}_{\rm eff}^{\bar{e}{\rm-}e}
 = 32 \p^3  \int \rd^6 x \sqrt{-g} \left( R^2 + {1\over 6 }
\na H \na H \right)\, .
\label{621b}
\ee
\item{}
In the ${\bar{o}{\rm-}o}$ sector we find the same result, but with
 an overall minus sign
depending on whether we consider type IIA or IIB ($\vep=1$ in IIA and $\vep=-1$ in IIB)
:
\be
{\cal I}_{\rm eff}^{\bar{o}{\rm-}o}
 =  32 \p^3 \vep \int \rd^6 x \sqrt{-g}
\left(  R^2 + {1\over 6}  \na H \na H\right)\, .
\label{628b}
\ee
Therefore,  one-loop string corrections generate $\Rr^2$ and
$\na H \na H$ terms in the effective action of type IIA
superstring
on K3, while no such terms appear in the type IIB superstring.
\item{}
The CP-odd sectors ${\bar{e}{\rm-}o}$ and ${\bar{o}{\rm-}e}$ again lead
to
the same vertices up to a sign depending on type IIA, B but also
on the nature of the particles involved. This leaves
\bs
\be
{\cal I}_{\rm eff, \ IIA}^{\rm CP-odd} =
 32\p^3 \int \rd^6 x \sqrt{-g}\,
{1 \over 2}\, ( B\w R \w R + B\w \na H \w \na H )\, ,
\ee
\be
{\cal I}_{\rm eff, \ IIB}^{\rm CP-odd} =
 - 32\p^3 \int \rd^6 x \sqrt{-g}\,
{1 \over 6}\,   H \w H \w R \, .
\ee
\label{637}
\es
\end{itemize}
Summarizing, we can put the results (\ref{621b}), (\ref{628b})
for the CP-even
terms and (\ref{637}) for the CP-odd terms together, and we record
the one-loop four-derivative terms in the six-dimensional effective
action for type IIA and IIB:
\bs
\be
{\cal I}_{\rm eff, \  IIA} = {1\over \pi}\int \rd^6 x \sqrt{-g}
\left( 2  R^2  + {1\over 3} \na H \na H + {1\over 2}
 B \wedge ( R \wedge R + \na H \wedge \na H) \right)\, ,
\label{638}
\ee
\be
{\cal I}_{\rm eff, \ IIB} =  - {1\over 6\pi}\int \rd^6 x \sqrt{-g}
              \,  \, H \wedge  H \wedge R\, ,
\label{639}
\ee
\es
The CP-odd term $B\w tr(R\w R)$ was first calculated in \cite{vw1}.
There, it was also explained how it can be obtained from the 
analogous ten-dimensional
term $B\w \Rr^4$ by reducing on K3.

As a check note that the type IIA theory should be invariant
under a combined space-time ($P$) and world-sheet parity ($\O$).
Since the Levi--Civita $\e$ tensor changes sign under $P$ while the $B$
field
changes sign under $\O$, we verify the correct invariance under $P\O$.
On the other hand, the type IIB theory is correctly invariant under the
world-sheet parity $\O$, since the interactions contain an even number
of antisymmetric tensor fields.

We should stress here that these thresholds, although they were
computed at
the $T^4/Z_2$ orbifold point of K3 are valid for any value of the
K3 moduli.
The reason is that the threshold is proportional to the elliptic genus
of K3 (which in this case is equal to the K3 Euler number) and thus
is moduli-independent.
It can also be seen directly in the $T^4/Z_2$ calculation as follows.
The result is obviously independent of the ($4,4$) orbifold moduli.
All the other moduli have vertex operators that are proportional to the
twist fields
of the orbifold. The correlator of three gravitons or antisymmetric
tensors
and one of the extra moduli is identically zero, since the symmetry
changes the sign of twist fields. Thus, the derivatives of the
threshold with respect to the extra moduli are zero.

\setcounter{equation}{0}
\subsection{One-loop gravitational corrections in four-dimensions
\label{4d}}
\label{susy4D}

Further compactification of six-dimensional $N=2$ type IIA, B string
theory
on a two-torus yields $N=4$ string theories in four dimensions.
Six-dimensional duality between heterotic string on $T^4$ and type IIA
string on K3 is expected to
descend to a duality between the corresponding four-dimensional $N=4$
compactified
theories.\footnote{For a derivation of the explicit map see \cite{book,lec}.}. 
The thresholds here will depend on the two-torus moduli $T,U$.
We will be interested in computing the moduli dependence of
the four-derivative terms involving the graviton, antisymmetric tensor
and dilaton, more generally called gravitational thresholds. The
terms of interest are therefore:
\bea
{\cal I}_{\rm eff}
= {2\over 3}\int \rd^4 x \sqrt{-g} & \Big( &\! \! \! \!
\D_{\rm gr} (T,U) R_{\m\n\r\s} R^{\m\n\r\s} +
\Theta_{\rm gr} (T,U) \e^{\m\n\r\s}_{\phl} R_{\m\n\a\b}^{\phl}
R_{\r\s}^{\hphantom{\r\s}\a\b}
\cr
&+& \! \! \! \!
\D_{\rm as} (T,U) \na_\m H_{\n\r\s}\na^\m H^{\n\r\s} +
\Theta_{\rm as} (T,U) \e^{\m\n\r\s}_{\phl}
\na_\m^{\phl}H_{\n\a\b}^{\phl} \na_{\r}^{\phl}
H_\s^{\hphantom{\r}\a\b} \cr
&+&\! \! \! \!
\D_{\rm dil} (T,U) \na_\m \na_\n \Fi \na^\m \na^\n \Fi +
\Theta_{\rm dil-as}(T,U) \e^{\m\n\r\s} \na_\m \na_\a \Fi \na^\a
H_{\n\r\s}
\cr
&+&\! \! \! \!
\Theta_{\rm gr-as} (T,U) \e^{\m\n\r\s}_{\phl} R_{\m\n\a\b}^{\phl}
\na_{\r}^{\phl} H_\s^{\hphantom{\r}\a\b}\,
\Big)\, .
\label{eac}
\eea
Again, we will use a short-hand notation for each term appearing in
the above expression:
$R^2$, $R\w R$, $\na H \na H$,
$\na H 
 \na H$, $\na\na\Fi\na\na\Fi$, $\na\na\Fi\w\na H$, $R \w \na
H$. Note
that there is no non-vanishing on-shell  $ R H $-coupling between one
graviton and one two-form, nor any $\na\na\Phi\w\na\na\Phi$ or
$\na\na\Phi\w R$ couplings.

The thresholds in  Eq.  (\ref{eac}), as advocated earlier, are  expressible
in terms of the $\lambda ^4$
helicity supertrace \cite{6} and, as such, will receive contributions from
1/2-BPS states only.

By direct calculation \cite{HM,6} we can extract the derivatives of the 
thresholds with respect to
the torus moduli.
They are expressible as integrals over the two-torus partition function.
Moreover each of the thresholds depends only on one complex modulus but not the other.
For example

\bs
\label{tdc}
\be
\mbox{type IIA:}\, \cases{
\pa_T \D_{\rm gr} =  \int_{\cal F} {\rd^2 \t \over \t_2}  \pa_T B_4
\cr
\pa_U \D_{\rm gr} = 0
\cr}
\label{da1}
\ee
\be
\mbox{type IIB:}\, \cases{
\pa_T \D_{\rm gr} = 0
\cr
\pa_U \D_{\rm gr} = \int_{\cal F} {\rd^2 \t \over \t_2}  \pa_U  B_4\, .
\cr}
\label{db1 tdc}
\ee
\es
We recover in this way the well-known result that $\Delta_{\rm
gr}$
only depends on the K\"{a}hler moduli $T$ and not on the
complex-structure
moduli $U$ in type IIA, while the reverse is true in type IIB
\cite{n=2t1}.
Similar interferences occur for all thresholds and yield the
following moduli dependences:
\bs
\be
{\rm IIA}: \; \;
\Delta_{\rm gr}(T)\sp
\Delta_{\rm as}(U)\sp
\Delta_{\rm dil}(U)\sp
\Theta_{\rm gr}(T)\sp
\Theta_{\rm as}(T)\sp
\Theta_{\rm gr-as}(U)\sp
\Theta_{\rm dil-as}(U)\, ,
\ee
\be
{\rm IIB}: \; \;
\Delta_{\rm gr}(U)\sp
\Delta_{\rm as}(T)\sp
\Delta_{\rm dil}(T)\sp
\Theta_{\rm gr}(U)\sp
\Theta_{\rm as}(U)\sp
\Theta_{\rm gr-as}(T)\sp
\Theta_{\rm dil-as}(T)\, .
\ee
\es
The dependence of $\Delta_{\rm gr}(T)$ is consistent with
our argument that the $\Rr^2$ term does not get corrections
beyond one loop.
However, there exists a subgroup of $SO(6,N_V,Z)$ that exchanges
the (type IIA) $U$-modulus with the dilaton $S$-modulus,
so that $SO(6,N_V,Z)$ duality implies that $\Delta_{\rm as},\Delta_{\rm
dil},
\Theta_{\rm gr-as},\Theta_{\rm dil-as}$ are also $S$-dependent, i.e.
are
perturbatively and non-perturbatively corrected. The loophole
in the argument of Section 2 is that, for these couplings, the
world-sheet instantons of the type IIB string are non-zero
(since they depend on the type IIB $T$-modulus), and therefore
the $(p,q)$ D 1-branes do contribute to instanton corrections.
{}From now on we shall restrict ourselves to $\Rr^2$ thresholds,
for which the type II one-loop result is exact.

The helicity supertrace $B_4$ entering in the threshold
can be readily computed  with the result:
\be
B_4 = 36~\Ga_{2,2}\, .
\label{b41}
\ee
We insert $B_4$ into Eq.  (\ref{da1}) and use the
fundamental-domain integral \cite{DKL} to obtain the $\Rr^2$
thresholds \cite{HM}:
\bs
\be
\mbox{type IIA:}\; \;  \D_{\rm gr} (T) =  -36 \log \left(
T_2 \left|
\et(T)\right|^4 \right)
 + {\rm const.}\, ,
\label{789} \ee
\be
\mbox{type IIB:}\; \;  \D_{\rm gr} (U) = -36  \log \left(
U_2 \left|
\et(U)\right|^4 \right)
 + {\rm const.}\, ,
\ee
\es
where the constant is undetermined.
Note that the one-loop thresholds are respectively invariant under
SL(2,\Z)$_T$ and
SL(2,\Z)$_U$, as they should.

\subsection{CP-odd couplings and  holomorphic anomalies}
\setcounter{equation}{0}

Moving on to the CP-odd couplings and focusing on the IIA case for
definiteness,
we find \cite{6}
\be
\partial_T\Theta_{\rm gr}=- {9i\over 2\pi^2} \partial_T\log \left(T_2\left|
\et(T)\right|^4\right)\sp
\partial_{\bT}\Theta_{\rm gr}={9i\over 2\pi^2}\partial_{\bT}\log\left(T_2\left|
\et(T)\right|^4\right)\, .
\label{640}
\ee
Would the non-harmonic $T_2$ term be absent, those two equations
could be easily integrated and would give
\be
\Theta_{\rm gr}(T) = {9\over 2\pi^2}\Im \log \eta^4(T)\, .
\ee
However, in the presence of the $T_2$ term the notation
$\pa_T \Theta$ and $\pa_{\bar{T}} \Theta$
for CP-odd couplings between two gravitons
and one modulus no longer makes sense.
This non-integrability of CP-odd couplings
has already been encountered before \cite{a2}. This
problem can be evaded
simply by rewriting the CP-odd coupling as
\be
{\cal I}^{\rm CP-odd}_{\rm gr}=\int\O \wedge  (Z_T \rd T + Z_{\bT} \rd
\bT)\, ,
\label{641}
\ee
where $\O$ is the gravitational Chern--Simons three-form, such that
$\rd \O=R\w R$.
In the special case $Z_T=\partial_{T}\Theta(T,\bT)$,
$Z_{\bT}=\partial_{\bT}\Theta(T,\bT)$, one retrieves by partial
integration the usual integrable CP-odd coupling. In the case at hand,
\be
Z_T=-{9i\over 2\pi^2} \partial_T\log \left(T_2\left|
\et(T)\right|^4\right)\sp
Z_{\bT}={9i\over 2\pi^2}\partial_{\bT}\log\left(T_2\left|
\et(T)\right|^4\right)\, .
\label{6422}
\ee
We can take advantage of the special structure of Eq.  (\ref{6422}) and
rewrite Eq.  (\ref{641}) as
\be
{\cal I}^{\rm CP-odd}_{\rm as}={3\over \pi^2} \p\int
\left({\Im}\left(\log\eta^4(T)\right)R \w R-
{1\over T_2} \, \Omega \w \rd T_1 \right)\, .
\label{643p}
\ee
In the decompactification limit $T_2\to \infty$, only the first
term survives and we obtain
\be
{\cal I}^{\rm CP-odd}_{\rm gr}={3\over \pi^2}\int\left({\pi\over 3}\,
T_1 \, R\w R
+{O}(1/ T_2)\right)\, .
\label{644}
\ee
This reproduces the six-dimensional type IIA result (\ref{638}).

\subsection{From the type-II to the heterotic string}
\setcounter{equation}{0}

The type II theory compactified on K3$\times T^2$ is dual to the heterotic 
string compactified on $T^6$.
The duality map exchanges $S$ and $T$, where $S$ is the
axion--dilaton multiplet, sitting in the gravitational multiplet
on the heterotic side.
It also acts by electric-magnetic duality on the two gauge fields coming from the 
antisymmetric tensor on the $T^2$ \cite{lec,book}.

Contrary to the type-II theory, the heterotic string theory possesses
a tree-level $\Rr^2$ coupling\footnote{This coupling exists in the axion-dilaton frame.
In the frame with the antisymmetric tensor this coupling becomes the gravitational Chern-Simons correction to the antisymmetric tensor field strength.}
required for anomaly cancellation
through the Green--Schwarz mechanism, together with an $\Rr^2$ coupling
required for supersymmetry.
The world-sheet fermions now have 8 zero-modes, so that the one-loop
three-particle amplitude vanishes (in even spin structure, one would
need four fermionic contractions to have a non-vanishing result
after spin-structure summation). In particular, we conclude that
there is no one-loop correction  to tree-level $\Rr^2$ coupling.

We can therefore translate the type IIA result
(\ref{789}) for the heterotic string on $T^6$:
\be
%\he{22} : \;\;   
\D_{\rm gr}  (S) = - 36 \log \left( S_2 \left|\et (S)
\right|^4 \right)=
12\pi S_2-36\log(S_2)+\label{hsd}
\ee
$$
+72\sum_{N=1}^{\infty}\left(\sum_{p|N}{1\over p}\right)\left[e^{2\pi iNS}+
e^{-2\pi i N\bar S}\right]
$$

The ${S_2 \ra \infty}$ heterotic weak-coupling limit exhibits
the tree-level $\Rr\wedge Rr$ coupling together with a non-perturbative
logarithmic divergence. Such a logarithmic divergence 
is also present in other instances~\cite{kp1}.
The full threshold is manifestly invariant under SL(2,\Z)$_S$,
and could in fact be inferred from SL(2,\Z)$_S$ completion
of the tree-level result. The exponentially suppressed terms
in Eq.  (\ref{hsd}) can be identified \cite{HM} with the instanton
contributions of the neutral heterotic NS5-brane wrapped on $T^6$,
the only instanton configuration preserving half the space-time supersymmetry
that can possibly occur in four-dimensional heterotic string.

The same mapping can be executed for the CP-odd $R\w R$ coupling:
\be
{\cal I}^{\rm CP-odd}_{\rm gr}=18  \int \left(\Im
\left(\log\eta^4(S)\right)
R\w R-{1\over S_2}\, \O \w  \rd S_1 \right)\, .
\label{643h}
\ee
There, however, in addition to the tree-level term and instead
of the logarithmic divergence, we find a coupling between
the axion and the gravitational Chern--Simons form.
Dualizing the axion into a two-form and keeping track
of the powers of the heterotic coupling $S_2$, this translates
into a {\it one-loop} coupling $H_{\m\n\r} \Omega^{\m\n\r}$
between one two-form and two gravitons, excluded by
a one-loop heterotic calculation. Happily enough,
the Chern--Simons form is co-closed, so that this coupling is a
total derivative and does not contribute to matrix elements.

\subsection{NS5-brane instantons}
\setcounter{equation}{0}

The heterotic NS5-brane is a BPS 5-brane that 
breaks half of the supersymmetry \cite{chs}. 
Its long-range fields in the transverse space include four-dimensional instanton configurations.
In the heterotic theory, the zero mode
fluctuation spectrum for a thick five-brane is composed of hypermultiplets.
It was further shown using heterotic/type-I duality that in the limit of zero thickness
(zero instanton size) there is an SU(2) gauge symmetry restored \cite{zerosize}.
The most important for us terms of its 
 world-volume action are CP-even Nambu-Goto volume term as well as the the CP-odd coupling to the dual of the antisymmetric tensor. The other terms involving the gauge fields as well as the charged hypermultiplets are not excited in a supersymmetric (BPS) configuration
 and we will ignore them.
 Thus,
\be
S_{\rm 5-brane}=T_5\int d^6 \xi ~e^{-\Phi}\sqrt{\det \hat G}+iT_5\int d^6\xi ~\tilde B_{012345}+\cdots
\label{5-b}\ee 
The induced fields are defined as 
\be
\hat G_{ab}=G_{\mu\nu}{\partial x^{\mu}\over \partial \xi^a}{\partial x^{\nu}\over \partial \xi^b}
\ee
and similarly for the six-form $\tilde B_{\mu_1\cdots,\mu_6}$ that is the dual of $B_{\mu\nu}$ in ten dimensions.
$x^{\mu}$ are coordinates in ten-dimensional space-time whereas $\xi^a$ are the coordinates of the six-dimensional world-volume.
The dots in (\ref{5-b}) stand for interactions that are not relevant for our analysis.
The tension $T_5$ can be obtained by saturating the Nepometchie-Teitelboim
quantization condition \cite{peros,nt} (the analogue of the Dirac quantization condition for branes) which in ten dimensions reads
\be
T_p T_{6-p}={2\pi n \over 2\kappa_{10}^2}
\label{nep}\ee
The electric dual of the heterotic NS5-brane is the perturbative heterotic string with tension $T_1=1/(2\pi \alpha')$.
Using $2\kappa_{10}^2=(2\pi)^7\alpha'^4$ and (\ref{nep}) for n=1, we obtain the  NS5-brane tension
\be
T_5={1\over (2\pi)^5\alpha'^3}
\ee
Remember that the full tension is $e^{-\Phi}T_5=T_5/g_s^2$ 
where $g_s$ is the ten-dimensional heterotic string coupling.

We can use the definition of the six-form $\tilde B$,

\be
dB=e^{-\Phi} ~^* d\tilde B\;\;\to\;\;(\partial_{\mu_1}\tilde B_{\mu_2\cdots\mu_7}+{\rm cyclic})={1\over 3!}{{\epsilon_{\mu_1\cdots\mu_7}}^{\mu_8\mu_9\mu_{10}}\over \det ~ G}~e^{-\Phi}~(\partial_{\mu_8}B_{\mu_9\mu_{10}}+{\rm cyclic})
\ee
and the definition of the four-dimensional axion
\be
e^{-\phi_4}H_{\m\n\rho}={{\e_{\m\n\rho}}^{\s}\over \sqrt{-g}}\partial_{\s}a
\ee
to show that 
when the ten-dimensional $B_{\mu\nu}$ has only four-dimen\-si\-o\-nal (transverse) dependence
then 
\be
{1\over 6!}\epsilon^{\mu_1\cdots\mu_6}\tilde B_{\mu_1\cdots\mu_6}=a
\ee
We have suppressed above an overall constant that cannot be determined from the duality transformation. This is necessary in order to match the instanton action.
 
We  must now consider the NS5-brane rotated to Euclidean space. 
Moreover, in order for it to have a finite action, its (Euclidean) world-volume must wrap
(supersymmetrically) around $T^6$.
We will now calculate $e^{-S_{Class}}$ and show that it has the form expected from duality in (\ref{hsd}).

We will take the world-volume of the NS5-brane to be also a six-torus.
The supersymmetric map is then
\be
X^a={M^a}_b\xi^b
\ee
where ${M^a}_b$ is an integer valued matrix.
The volume of the target $T^6$ is  $ (2\pi)^6\alpha'^3 V_6$. 
It is straightforward to evaluate
\be
\int d^6\xi \sqrt{\det \hat G}= (2\pi)^6\alpha'^3 V_6|det ~M|
\ee
$N=det ~M$ is the ``winding number", that tells us how many times the brane is wrapping around the torus.
We also have
\be
\int d^6\xi ~\tilde B_{012345}=(2\pi)^6\alpha'^3 ~a ~det~M
\label{axi}\ee
for constant axion.
Putting everything together we obtain
\be
S_{class}=T_5\int d^6 \xi ~e^{-\Phi}\sqrt{\det \hat G}+iT_5\int d^6\xi ~\tilde B_{012345}=2\pi |N| {V_6\over g_s^2}+2\pi Ni a=
\ee
$$
=2\pi |N|{1\over g_4^2}+2\pi iNa=
2\pi (|N| ~S_2+iNa)
$$
where in the last equality we have introduced the four-dimensional (dimensionless) heterotic string coupling as the imaginary part of the complex S field.
Thus, for positive N we obtain the instanton factor
\be
e^{-S_{class}}=e^{2\pi i NS}
\ee
which is holomorphic in $S$.
For N negative we obtain the anti-instanton contribution instead 
\be
e^{-S_{class}}=e^{-2\pi i N\bar S}
\ee
These instanton correction factors have exactly the form predicted by duality
in (\ref{hsd}).

Duality predicts that the determinant is $\sim \sum_{p|N}{1\over p}$, while there are 
no further corrections.
It is hard to see how one could reproduce this determinant from a NS5-brane calculation.
If the brane is free to wrap any possible way around $T^6$ without any other factor,
then we should be gauge-fixing the SL(6,Z) world-sheet symmetry acting on the left on M
and sum on the left-over entries.
This does not reproduce the result. Rather, it seems that the brane wraps in a unique
way on a $T^4$ subtorus, and then freely (modulo SL(2)) on the left-over $T^2$.
It would be very interesting to calculate the determinant from first principles.
A promising approach would be to do the calculation in the type-I dual picture in which case we will be dealing with the type-I D5-brane.

There are further N=4 D=4 type-II ground-states that are dual to
heterotic string ground-states. 
Some of them are not left-right symmetric so that they
do not have a direct geometrical interpretation.
Moreover, the Montonen-Olive duality relevant in the heterotic side
corresponds to proper subgroups of SL(2,$\Z)_S$.
Similar tests for these extended dualities (corresponding to calculating the $\Rr^2$ and other thresholds) have been carried out in
\cite{6,grego}.

\subsection{Absence of d=4 instanton corrections for $tr\F^4$ in the N=4 heterotic 
theory}
\setcounter{equation}{0}

We had argued that the first dimension where non-perturbative effects are expected 
to modify BPS-saturated amplitudes in the heterotic string is D=4.
There is however already a field theory result \cite{ds2} that claims that non-perturbative effects cannot renormalize $\F^4$ couplings in N=4 four-dimensional 
super Yang-Mills
theory.
There, the full result is one-loop only.
We will argue here that the same is true when gravity is included for the $tr\F^4$ terms
(instanton effects cannot be excluded for the graviphoton $\F^4$ terms).
 
Heterotic/type IIA duality in six dimensions implies that the one loop $tr\F^4$ terms
in the heterotic theory should be generated at tree level in the type-II theory.
Moreover it is easy to argue that there cannot be any D-instanton corrections to such terms in six dimensions in the type IIA theory.
The reason is that the potential instantons are due to D0, and D2 branes and these need
supersymmetric one- and three-cycles inside K3 and there is none.  
Let us consider further compactification on a two-torus.
The tree-level result will remain independent from the volume $T_2 $ of the two torus.
Moreover, now there are potential non-perturbative corrections due to D0,D2,D4 and NS5
branes.
All the D-branes cannot wrap supersymmetrically the whole of $T^2$.
They wrap one-cycles, the contribution is proportional to the  length $~\sqrt{T_2}$ but this
makes $1/g_6\to 1/g_4$ up to the complex structure modulus.
Consequently the threshold due to these is independent of $T_2$.
The same is true for the NS5-induced threshold, since there the wrapping of NS5 around K3$\times T^2$ would produce $1/g_4^2=T_2/g_6^2$.
The upshot of the above is that the full $\F^4$ threshold in four-dimensional 
type II theory on K3$\times T^2$ is dependent only on O(6,22) moduli, and is independent
on the SL(2)/U(1) modulus, T.

The above automatically implies that the heterotic results is independent of $S$ so that
in particular there are no instanton corrections in d=4 to $tr\F^4$ terms in the 
heterotic string \cite{kop}.

\section{N=8, $\Rr^4$ couplings and D-brane instantons in type II string theory}
\setcounter{equation}{0}

Another context where our knowledge on stringy instanton corrections has improved 
considerably recently is that of the $\Rr^4$ thresholds in the toroidal compactifications of type-II string theory with maximal supersymmetry.
We will be brief here since there are already three good recent reviews of several aspects of  the problem at hand \cite{piophd,op,mbg}.
There, various aspects of the problem are treated in more detail.
Here we will only provide an overview.

In ten dimensions, for maximal N=8 supersymmetry, the two $\Rr^4$ invariants are both
1/2-BPS-saturated (as discussed in section 3.4).
They get contributions from tree level and one-loop only.
A detailed discussion of the tree and one-loop contributions in the type-IIA,B 
strings can be found in \cite{kp1}. In \cite{ie} a two-loop calculation of the four graviton amplitude was performed. This implies a potential two-loop contribution 
for the $\Rr^4$ threshold \cite{ie2}, which is not obviously zero. 
There are independent
arguments though that indicate that the two and higher loop contributions 
would vanish \cite{berk1,pio,grs}.
The way out would be that the two-loop integrand \cite{ie2} is a total 
derivative on moduli space.
 
In the ten-dimensional type-IIA case there are no potential instanton corrections.
The lowest order D-brane here, is a D0 brane which needs a one-dimensional compact manifold in order to provide a finite instanton correction.
In ten-dimensional type-IIB theory, however, we have D(-1) instantons, 
that break half of the space-time supersymmetry and are thus, just right to give non-perturbative corrections  to the $\Rr^4$ threshold.
In fact, such non-perturbative corrections are required by the conjectured SL(2,\Z) invariance of the ten-dimensional type-IIB theory.
The non-perturbative threshold was first conjectured \cite{GG2} to be given by 
the SL(2,\Z), weight-3/2 Eisenstein series which was SL(2,\Z) invariant and matched perturbation theory.
Afterwards, this conjecture was found to be a consequence of 
heterotic/type II duality \cite{anto}, 
and the SL(2,\R) structure of N=2 supersymmetry \cite{berk1,pio,grs}.

Another insight was provided in \cite{GV} where the threshold was calculated by using 
an eleven-dimensional perspective.
Upon compactification in nine dimensions, we have two complementary pictures.
In the type-IIB theory we still expect only D(-1) instanton contributions which, 
since they are localized in space-time, are essentially the same as in ten 
(up to trivial compactification factors). 
The next type-IIB brane namely the D1-brane needs a 
two-dimensional compact manifold to provide an instanton and can start 
contributing in eight or lower dimensions, but not in nine.
On the other hand, in the type-IIA theory, the D0 branes can now wrap their world-lines
around the $S^1$ and provide instanton contributions to the $\Rr^4$ threshold.
Moreover, the type IIA and IIB thresholds are mapped to one-another by the $R\to 1/R$ standard T-duality relating the type-IIA,B vacua in nine dimensions.
Now D0-branes are essentially the Kaluza-Klein modes of the supergraviton 
of the eleven-dimensional M-theory.
This gives an intuitive explanation why the full threshold can be calculated 
in nine-dimensions from a one-loop amplitude in eleven-dimensional supergravity
(there is also a subtlety: the power divergence of this amplitude must be cut-off 
by hand \cite{GV}).
Moreover, in a generic compactification of the type-II theory on a $T^n$ torus to 10-n dimensions, the eleven-dimensional one-loop supergraviton amplitude is giving the D0-instanton contribution to the threshold.
However in $d<9$ dimensions there are further non-perturbative contributions.

The problem at hand is: can we calculate the $\Rr^4$ threshold at various 
lower-dimensional ground-states?
The motivation for this is manyfold:
The effective $\Rr^4$ terms and their supersymmetric partners \cite{kpar1,grs}
are the leading $\alpha'$-corrections to the supergravity effective action (with maximal supersymmetry)
and they are useful for checking the departure from field theory in many contexts.
For example, stringy corrections to black-hole entropy can be associated with 
these effective terms \cite{msw}.
Such CP-even terms are related by supersymmetry to CP-odd terms that are crucial
in anomaly (inflow) arguments \cite{anoi}.
Thus, their (non-perturbative) corrections are also of importance.
   Finally, we need to understand the quantitative rules of D-instanton calculus, which can be useful in other more complicated situations.

The essential approach for dealing with lower dimensions is a 
tumbling-down process by jumping
alternatively between type-IIA and type-IIB language.
We have already argued that in nine-dimensions by T-duality we have understood
the non-perturbative corrections as Euclidean D0-branes wrapping around the $S^1$.
Going to d=8 by compactifying the type-IIA theory on $T^2$, we can still calculate the D0-contribution: the only difference here is that D0-branes can now wrap on either cycle of the $T^2$. Moreover since the D2-branes need a three-manifold to give instanton corrections, the D0-brane result captures all non-perturbative corrections in d=8.
Now, by a T-duality we translate the threshold in type-IIB language.
Here we expect both D(-1) and D1-instanton corrections.
Since the D(-1)-contributions are already understood in d=10, knowledge of the full result allows us to dissentagle the D1-contributions.
Thus, at the end, in d=8 we know all D(-1), D0 and D1 rules.
Going now to d=7 by compactifying the theory on $T^3$, we can now do the full calculation in the type-IIB language. The reason is that all non-perturbative contributions
come from the D(-1) and D1 branes (the D3 brane needs a four-manifold to contribute).
By T-dualizing we can learn the D2-brane contribution in the type-IIA theory which contributes in d=7.
It is obvious that by bouncing back and forth using T-duality between type-IIA and 
type IIB language we will be able to work out all Dp-brane instanton rules as well
as the $\Rr^4$ threshold in lower dimensions.

There are other possibilities though.
In \cite{kp1}, the focus has first been in d=8 type-IIB theory.
There, we have a one-loop contribution, which is given by the integral of the toroidal partition function on the fundamental domain of the torus modulus.
This integral is well known \cite{DKL} and gives a ``degenerate" contribution (depending on the complex structure U of the torus), and a non-degenerate contribution that contains
terms that are exponentially suppressed as functions of the torus volume $T_2$.
These are contributions of "fundamental string instantons", where the string wraps its
world-sheet around the two-torus.
The SL(2,\Z) symmetry of the theory implies that apart from the fundamental (1,0) string we also have (p,q) strings. The (0,1) string is the D1-string, and the (p,q) strings
can be thought of as bound states at threshold \cite{boun}.
Using SL(2,\Z) to calculate the effective torus volume, the full threshold can be
written as a sum of the D(-1) instanton contribution as well as a sum over the 
instanton contributions of the (p,q) strings.
This almost captures the full threshold. The subtlety arises from the 
logarithmic infrared divergence of the threshold in eight dimensions.
An extra logarithm of the moduli needs to be added so that the threshold is invariant under the U-duality group SL(3,\Z)$\times$SL(2,\Z).
The final result is the sum of the weight-3/2 Eisenstein series for SL(3,\Z) and the 
weight-1 Eisenstein series for SL(2,\Z).
This can be independently understood from eleven dimensions.
The relevant contributions is that of supergravitons on $T^3$ that generate the SL(3,\Z) Eisenstein series as well as the M2-brane wrapped on $T^3$ that generates the SL(2,\Z)
Eisenstein series.

In seven dimensions, the U-duality group is SL(5,\Z).
A natural guess for the threshold is the weight-3/2 SL(5,\Z) Eisenstein series.
This turns out to be correct \cite{kp1} as it gives the correct one-loop threshold and
its non-perturbative piece agrees with the D(-1) and D1 instanton corrections that are expected in the seven-dimensional type-IIB theory.

Finally, there is the procedure inspired by the eleven-dimensional origin 
of the type-IIA theory \cite{kp2,neuc}.
The idea is the following: The supergraviton one-loop threshold of the eleven-dimensional
theory compactified on $T^n$, appropriately regularized, gives the D0-instanton contribution
of the type-II theory compactified on $T^{n-1}$.
Picking a direction in the torus, say the first, we can perform a T duality transformation.
This automatically gives as the amplitude for D1 branes in the 1-direction as well as 
D(-1) instantons.
However, the T-dualization has been done in a specific direction.
In order to obtain the most general D1+D(-1) instanton contribution on $T^n$ we need to "covariantize" the result in the eleven-dimensional sense.
This can be done, \cite{kp2}.
Continuing further with T-duality, we can obtain the D2-D0 instanton contribution,
the D3-D1-D(-1) instanton contribution etc.
Furthermore, such contributions (as well as the one-loop ones) 
can be neatly described as appropriate generalized Eisenstein series for the U-duality 
groups $E_{n(n)}(\Z)$ \cite{eisen}.
The precise conjecture is that the $\R^4$ threshold is proportional to the 
weight s=3/2 E$_d(d)$
Eisenstein series for the string representation (the {\bf 5} for SL(5,\Z), the {\bf 10}
for SO(5,5,\Z), the {\bf $\bar {27}$}  of E$_{6(6)}$, the {\bf 133} of E$_{7(7)}$ etc. 
\cite{op}).

This procedure was worked out in detail up to six-dimensional toroidal 
compactifications of type II theory.
The the six-dimensional $\Rr^4$ threshold seems to contain $e^{1/g_s^2}$
contributions that are characteristic of NS5-instantons but 
which,  on the other hand, cannot contribute in six dimensions.
Their contribution though is not uniform \cite{eisen}.
The answer to this puzzle is so far unknown.

\section{Summary and Open Problems}

We have given a survey of BPS-saturated terms in extended supersymmetric theories.
Such terms:

(1) obtain perturbative corrections from BPS states only.

(2) The perturbative corrections appear at a single order of perturbation
theory, usually
at one-loop.

(3) They satisfy "holomorphicity constraints".

(4) They obtain instanton corrections from "BPS-instantons" (instanton
configurations that
preserve some fraction of the original supersymmetry).

(5) If there exists an off-shell formulation they can be easily constructed.

Due to the properties cited above, they are important in testing the consistency 
of non-perturbative dualities in supersymmetric theories.
They are also central in understanding the detailed rules of non-perturbative dualities
and in particular those of instanton calculus.

We have given an extended example of such techniques here: 
we analysed in some detail the issue of heterotic/type-I duality. 
The two theories are dual in ten dimensions.
The relevant BPS-saturated couplings that obtain non-trivial corrections are $\F^4$
and $\Rr^4$ type couplings. Their coefficients match properly in ten dimensions. 
Upon compactification on a circle, it turns out that the thresholds still match 
in perturbation theory (up to contact terms that we will return to later).
This implies that no soliton loops are necessary for agreement.
Moreover, in nine-dimensions no instanton contributions are expected in either theory.
 
Compactifying to eight dimensions, the heterotic thresholds are still one-loop.
However, now instanton corrections are expected on the type-I side due to the Euclidean
D1-brane wrapping around the two-torus.
This was confirmed and using duality we derived the relevant instanton sum.
The summation rules in the type-I side were also elucidated, and are 
in accord with what is expected from a matrix theory setup.
Moreover it can be checked that the picture works also in less than eight dimensions.

There has been more work in this direction confirming the picture above, and testing
further the duality as well as instanton calculations, \cite{fs,van,typeI3}.
Perturbative corrections around the instantons have been computed 
\cite{typeI4}.
Since the eight-dimensional heterotic string is expected to be dual to F-theory on K3, this suggested a geometric way of understanding $\F^4$ thresholds in the F-theory compactification. Such an understanding has been pursued in \cite{FF4} where the thresholds were written in terms data on multiple K3 surfaces. 

There are, however, several open problems.
As we saw, the heterotic threshold and duality implies that there should be extra contact
terms on the type-I side that seem to correspond to higher orders of type-I perturbation
theory ($\chi=-1,-2$).
Moreover, the $\Rr^2$ threshold on the annulus in the type-I theory, turns out to be 
non-zero \cite{fs} and depends on the complex structure modulus.
A similar situation (Planck-mass one-loop renormalization in type-I and heterotic
unmatched contact term ) occurred in N=2 heterotic/type-II compactifications to four
dimensions \cite{ ABFPT}.
There, the duality map had to be modified and accounted for the discrepancy.
It is not obvious what the resolution is in our case.

Another open problem is the application of the instanton rules derived here
to the type-I D5-brane. This will generate new instanton corrections in four dimensions for $\Rr^4$ terms or in three-dimensions
for the $\F^4$ terms. Moreover, via duality they will be related to NS5-brane instanton corrections on the heterotic side\footnote{Some as yet unpublished work in this direction was  reported to me by K. Narain.}.

There should also be an infinite series of BPS-saturated terms sensitive to 1/2 BPS
multiplets in N=4 supersymmetry. As mentioned earlier, their type-II 
thresholds should be related to topological amplitudes of the K3 $\s$-model, 
\cite{BV}. Moreover there should exist BPS-saturated terms (like $F^6$
terms that are sensitive to 1/4-BPS states.
The existence of such terms (namely $\Rr^4 {\cal T}^{4g-4}$, 
with {\cal T} the graviphoton field strength) is also suggested  in \cite{fs} where the appropriate thresholds have been computed in eight dimensions.
Thresholds of 1/4 BPS-saturated terms
are crucial in order to test the existence and spectrum of 1/4-BPS states 
due to string networks \cite{snet} 

In the type-I theory, the leading (disk) effective action is given by the Born-Infeld action that provides, if expanded, an infinite series of higher $\F^{2n}$ terms.
We have been analysing the $\F^4$ piece of that series. According to heterotic/type-I
duality the higher terms should come from two and higher loops in the heterotic theory.
This has never been checked.

We have further looked at another example where there is an interplay 
of duality and instantons. This was the case of heterotic/type IIA duality in six dimensions and its four-dimensional avatar.
On the type-II side, the $\Rr^2$ threshold is given by a one-loop result
and depends on the volume modulus $T$ of the $T^2$.
By duality $T\leftrightarrow S$ and this gives the tree level heterotic result
plus terms that as we have argued can be interpreted as NS5-brane instantons. 
This is another situation where duality implies some rules of instanton calculus,
which on the other hand do not look as natural as we had seen in the previous case.
It seems that the NS5-brane has to wrap once around $T^4$ and then in any possible way around the extra $T^2$.\footnote{A similar phenomenon seems to appear in the case of M2-brane instanton corrections to the $\Rr^4$ coupling in 
compactified type IIA theory.}
An natural interpretation of the instanton fluctuation determinant is lacking.

In six dimensions, there is  no extensive quantitative test of heterotic/type-II duality.
This would amount to a comparison of $\F^4$ and $\Rr^4$ terms on both sides.
As in the heterotic case, we do not expect non-perturbative corrections
on the type-II side for the $\F^4,\Rr^4,\Rr^2\F^2$ terms.
This can be seen as follows: the relevant D$p$-branes of the
ten-dimensional
IIA theory have $p=0,2,4,6,8$ with world-sheets being
$1,3,5,7,9$-dimensional.
To obtain an instanton contribution we need appropriate supersymmetric
cycles on K3 with dimension belonging to the list above.
It is known that there are  no such cycles.
Moreover, we also have the five-brane, which is magnetically coupled to
the NS-NS
antisymmetric tensor. Since its world-sheet is six-dimensional it can
only give
instanton corrections in $D<5$ dimensions.
Thus, in $D=6$, heterotic/type-II duality can be tested for the special
terms in perturbation theory.
The relevant objects on the
type-II side are the $N=4$ topological amplitudes defined in \cite{BV}.
The tree-level $F^4$
terms on the type-II side should match the one-loop corrections to such terms
on the heterotic side.
The infinite series of higher-loop  $\F^{2n}$ terms.
on the heterotic side   correspond here to an infinite sequence of tree level $\F^{2n}$ terms.

We can further compactify both theories on a circle to five dimensions.
There are still no non-perturbative corrections on the heterotic side.
In the type-II theory, we expect instanton corrections from the D2- and
D4-branes,  which are electrically (magnetically) charged under the
3-form.
The D2-brane can wrap around ${\cal S}^1$ and a supersymmetric two-cycle
of K3.
The D4-brane can wrap on ${\cal S}^1$ and the whole of K3.
These non-perturbative type-II corrections are expected to reproduce
the heterotic cross-terms coupling the (4,4) and the (1,1) lattice.
A more thorough investigation is needed, however.
 
Finally, in the last section we gave a brief account of 
$\Rr^4$ thresholds in toroidal compactifications of type-II string theory, and their relation to Dp-brane instantons.
We described  various ways of guessing or calculating the Dp-instanton
contributions.
Up to seven dimensions, the situation is simpler and the rules are well understood,
\cite{GG2,GV,kp1,kp2,po,neuc}. The situation in six dimensions 
has been worked out in \cite{kp2,eisen} and seems at present to indicate puzzling
$e^{1/g_s^2}$ non-perturbative terms (although the expansion is not under 
strict control).
There is certainly more to be learned in this direction.
Moreover, other higher derivative terms have been analysed in this context \cite{kpar2,berv}
and their non-perturbative thresholds have been linked to eleven dimensions \cite{kpar2,gopa}.

There is a further puzzle that was evident in the case of four-dimensional heterotic/type-II
duality for the $R^2$ terms, but it is in fact generic.
This is the problem of logarithmic moduli dependence due to logarithmic IR divergences.
In this particular context, the problem was as follows:
The one-loop type-II $\Rr^2$ threshold depends logarithmically on $T_2$ and this is due to the
(physical) IR-divergence.
This, upon duality implies for the heterotic threshold a term that is 
logarithmic in the string coupling.
The heterotic origin of such a term is at least obscure.
Another example is related to the $\Rr^4$ type-II couplings and 
discussed in \cite{kp1}.
In D=8, the $\Rr^4$ threshold is logarithmically IR divergent.
This is reflected by the appearance of moduli logarithms in the one-loop result.
Once we add the instanton contributions, the threshold is almost SL(3,\Z)-invariant
but not quite, \cite{kp1}. An extra logarithmic term has to be added in order 
to render it invariant.
There are more examples related to $\Rr^2$ thresholds in N=2 ground-states in four dimensions
\cite{r2anom}

\vskip 2.8cm
\centerline{\Large\bf Acknowledgements}
\vs .1cm
This research was partially supported by EEC grant
TMR-ERBFMRXCT96-0090.
I would like to thank my collaborators, C. Bachas, C. Fabre, A. Gregori,
C. Kounnas, N. Obers, B. Pioline and
P. Vanhove for collaboration, friendship, ideas and explanations.
I would also like to thank the organizers of the spring school for giving me the
opportunity to present these lectures. 
I am indebted to N. Obers and B. Pioline for a critical reading of the manuscript.

\ed